\documentclass[preprint,jcp,aip]{revtex4-1}
\usepackage[latin1]{inputenc}
\usepackage{amsmath}
\usepackage{amssymb}
\usepackage{graphicx}

\makeatletter

\providecommand{\tabularnewline}{\\}

\newcommand{\equationabb}{\text{Eq.\ }}

\newcommand{\equationsabb}{\text{Eqs.\ }}

\newcommand{\figureabb}{\text{Fig.\ }}

\newcommand{\figuresabb}{\text{Figs.\ }}

\newcommand{\tableabb}{\text{Tab.\ }}

\makeatother

\begin{document}

\title{Application of the Mixed Time-averaging Semiclassical Initial Value
Representation method to Complex Molecular Spectra}

\author{Max \surname{Buchholz}}

\affiliation{Institut für Theoretische Physik, Technische Universität Dresden,
01062 Dresden, Germany}

\affiliation{Dipartimento di Chimica, Università degli Studi di Milano, via C.
Golgi 19, 20133 Milano, Italy}

\author{Frank \surname{Grossmann}}

\affiliation{Institut für Theoretische Physik, Technische Universität Dresden,
01062 Dresden, Germany}

\author{Michele \surname{Ceotto}}

\affiliation{Dipartimento di Chimica, Università degli Studi di Milano, via C.
Golgi 19, 20133 Milano, Italy}
\email{michele.ceotto@unimi.it}

\begin{abstract}
The recently introduced mixed time-averaging semiclassical initial
value representation molecular dynamics method for spectroscopic calculations
{[}M. Buchholz, F. Grossmann, and M. Ceotto, J. Chem. Phys. 144, 094102
(2016){]} is applied to systems with up to 61 dimensions, ruled by
a condensed phase Caldeira-Leggett model potential. By calculating
the ground state as well as the first few excited states of the system
Morse oscillator, changes of both the harmonic frequency and the anharmonicity
are determined. The method faithfully reproduces blueshift and redshift
effects and the importance of the counter term, as previously suggested
by other methods. Differently from previous methods, the present semiclassical
method does not take advantage of the specific form of the potential
and it can represent a practical tool that opens the route to direct
\emph{ab initio} semiclassical simulation of condensed phase systems.
\end{abstract}
\maketitle

\section{Introduction}

In a recent publication \cite{Buchholz2016} it has been shown that
the ideas of time-averaging \cite{Elran1999,Kaledin2003} and of semiclassical
hybrid dynamics \cite{Grossmann2006} can be combined and lead to
an accurate description of molecular spectra of an anharmonic system
of interest in the presence of an environment. In the present work
the performance of the methodology is tested on systems with a much
larger total number of degrees of freedom than treated before and
we specifically answer the question under which condition a redshift
or a blueshift of the spectral line of the anharmonic system oscillator
are to be expected.

To this end, we employ a time-dependent approach to spectroscopy,
pioneered by Heller \cite{Heller1981_2}, based on the so-called semiclassical
initial value representation (SC-IVR) molecular dynamics introduced
by Miller \cite{Miller1970,Heller1975,Levit1977,Heller1981,Herman1984,Heller1991,Miller2001a}.
Ignited by seminal work of Kay in the early 90s \cite{Kay1994,Kay1994-1,Kay1994-2},
the literature has recently seen a flurry of activities in SC-IVRs,
and the Herman-Kluk (HK) approximation \cite{Herman1984,Grossmann1998}
has turned out to be the semiclassical method of choice of many authors.
\cite{Miller1998,Sun1999,Thoss2001,Wang2001,Wang2000,Gelabert2001,Yamamoto2002,Miller2005,Liu2007,Liu2007-1,Tao2011,Miller2006,Shalashilin2004,Kay2006,Sklarz2004,Harabati2007,Hochman2009,Herman1994,Makri1999,Zhang2004,Zhang2005,Shao2006,Pollak2007,MartinFierro2006,Walton1995,Walton1996,Brewer1997,Bonella2003,Bonella2005,Harabati2004,Grossmann1999,Viscondi2011,Antipov2015,Venkataraman2001,Nakamura2016,Kondorskiy2015,Tatchen2009,Huo2011,Matthew2017}
More recently, semiclassical molecular dynamics has been implemented
for on-the-fly simulations employing ab initio molecular dynamics
tools. \cite{Ceotto2009,Ceotto2009-1,Ceotto2011-1,Ceotto2013,Conte2013,Tatchen2009,Ianconescu2013,Wong2011,Wehrle2014,Wehrle2015,Zimmermann2012,Zimmermann2014}

The HK SC-IVR can, however, be only applied to a relatively small
number of coupled degrees of freedom. One route towards the description
of the spectra of larger systems is the addition of a time-averaging
filter to the phase space integration. \cite{Kaledin2003,Kaledin2003-1}
When the filter is fully exploited by taking long time-evolved classical
trajectories, the phase space integration numerical effort is reduced
by an order of magnitude. Further improvement in computational cost
to just a handful of classical trajectories is achieved by taking
into consideration that accurate eigenvalues can be obtained from
single trajectories when these are close in energy to the eigenvalues.
In fact, the trajectories whose energies are about the same as the
vibrational peaks' energies are contributing most to the spectroscopic
signal. This approach is called Multiple Coherent TA-SCIVR (MC-SCIVR
or MC-TA-SCIVR) and it has proved to be accurate for molecules such
as H$_2$O, CH$_4$, CH$_2$D$_2$ and NH$_3$. \cite{DiLiberto2016,Tamascelli2014,Conte2013,Ceotto2009}
The significant reduction in number of classical trajectories offered
by the MC-TA-SCIVR approach allowed to obtain quite accurate power
spectra of molecules using a direct ab initio dynamics simulation,
also called on-the-fly or direct ab initio semiclassical dynamics.
\cite{Gabas2017,DiLiberto2016,Tamascelli2014,Conte2013,Ceotto2013,Ceotto2011,Ceotto2011-1,Ceotto2010,Ceotto2009,Ceotto2009-1}
More recently, to beat the curse of dimensionality, a projection technique
for the MC-TA-SCIVR has been introduced. The new method is called
Divide-and-Conquer SCIVR (DC SCIVR) and it allows the calculation
of power spectra for high dimensional systems, such as a fullerene
buckyball molecule. \cite{Ceotto2017}. 

An alternative approach to reduce the number of classical trajectories
is Heller's thawed Gaussian wavepacket dynamics (TGWD) \cite{Heller1975},
where only a single Gaussian wavepacket with time-dependent width
is propagated. This numerically very cheap method is accurate only
for at most harmonic potentials, but it can be combined with the more
demanding HK method in the semiclassical hybrid dynamics formalism.
\cite{Grossmann2006} Propagating only few degrees of freedom on the
HK level, while using the simpler TGWD for the larger environmental
part, gives rise to an accurate yet efficient description for the
dynamics of systems with many degrees of freedom. We have recently
combined the hybrid idea with the time-averaging filter to arrive
at the mixed TA-SCIVR method (M-TA-SCIVR) that needs fewer trajectories
for convergence than a full TA-SCIVR treatment while being just as
accurate for the HK degrees of freedom. \cite{Buchholz2016} 

A model system that allows for an easy distinction of degrees of freedom
to be treated on the HK versus TGWD level is the one used by Caldeira
and Leggett (CL) in their seminal path integral studies of quantum
dissipation \cite{Caldeira1981}. This model with different analytical
forms of spectral density and cutoff has been widely used in different
branches of the quantum dynamics community to model system-bath interactions.
\cite{Pollak1986,Grossmann1995,Cao1997,Weiss1999,Gelabert2001,Tanimura2006,Bonfanti2012,Garashchuk2013}
Using a normal mode analysis, E. Pollak and coworkers have shown analytically
for a harmonic system with and without an additional cubic term that
the system frequency shift induced by the CL bath with an Ohmic spectral
density is always towards higher frequencies, i.e., a blueshift \cite{Pollak1988,Pollak1986}.
For the cubic system, another analytical study has also shown a blueshift
tendency for different bath spectral densities \cite{Williams1999}.
The same result has been obtained for a Morse oscillator coupled to
a CL bath \cite{Joutsuka2011}. On the other hand, arguing that experimental
results often report a redshift of the system frequency (for example
for iodine in rare gas matrices\cite{Karavitis2001,Karavitis2004}),
Georgievskii and Stuchebrukhov \cite{Georgievskii1990} have investigated
the influence of the CL counter term on a cubic system potential and
found that by omitting the counter term, both blueshift and redshift
are possible depending on bath parameters. We will employ a discretized
Ohmic spectral density, frequently applied in the CL model, and explicitly
treat the dynamics of the combined system, comprising the anharmonic
system of interest as well as the bilinearly coupled harmonic bath
modes. For up to ten bath degrees of freedom, this can still be done
on the Herman-Kluk level of the semiclassical description and serves
as a benchmark for our more approximate mixed semiclassical time-averaging
method that has to be used if up to 60 bath degrees of freedom are
taking part in the dynamics. Studying both cases of a non-resonant
as well as a resonant bath, we will show that both redshifts as well
as blueshifts are observed. 

The paper is organized in the following way: Section \ref{sec:mixed}
recalls the mixed time-averaging semiclassical method for the calculation
of molecular spectra. In Section \ref{sec:model}, we recapitulate
the CL model and discuss the discretization of the bath's spectral
density. In the central Section \ref{sec:results} results are first
given for ten bath oscillators and different levels of approximation.
The high quality of the results of our proposed approximation methodology
is thereby shown. Then a detailed study of the frequency shift of
the oscillator of interest in the presence of a substantial number
of bath degrees of freedom is performed. Conclusions are drawn and
an outlook is given in Section \ref{sec:conc}.

\section{Mixed Time-averaging Semiclassical Initial Value Representation}

\label{sec:mixed}

We first recapitulate the mixed time-averaging semiclassical initial
value approach to the calculation of molecular spectra. This method
has been introduced recently \cite{Buchholz2016} and combines the
semiclassical hybrid dynamics \cite{Grossmann2006} with time-averaging
\cite{Elran1999,Kaledin2003}.

The goal of the method is to calculate the power spectrum $I(E)$
of a given initial state $\left|\chi\right\rangle $ subject to a
Hamiltonian $\hat{H}$, 
\begin{align}
I(E)=\sum\limits _{n}\left|\left\langle \chi|\psi_{n}\right\rangle \right|^{2}\delta\left(E-E_{n}\right),\label{eq:IE_def}
\end{align}
where $E_{n}$ are the eigenenergies of interest and $\left|\psi_{n}\right\rangle $
are the corresponding eigenfunctions of $\hat{H}$. The spectrum can
be found from the system's dynamics by expressing the Delta function
as a Fourier integral. \equationabb (\ref{eq:IE_def}) then becomes
\begin{align}
I(E)=\frac{1}{2\pi\hbar}\int\limits _{-\infty}^{\infty}\text{d}t\ \text{e}^{\text{i}Et/\hbar}\left\langle \chi\middle|\text{e}^{-\text{i}\hat{H}t/\hbar}\middle|\chi\right\rangle \label{eq:IE_dyn}
\end{align}
The time evolution in \equationabb (\ref{eq:IE_dyn}) is calculated
semiclassically with the propagator by Herman and Kluk \cite{Herman1984},
\begin{align}
\text{e}^{-\text{i}\hat{H}t/\hbar}=\frac{1}{(2\pi\hbar)^{F}}\int\text{d}\mathbf{p}(0)\int\text{d}\mathbf{q}(0)\ C_{t}(\mathbf{p}(0),\mathbf{q}(0))\ \text{e}^{\text{i}S_{t}(\mathbf{p}(0),\mathbf{q}(0))/\hbar}\left|\mathbf{p}(t),\mathbf{q}(t)\right\rangle \left\langle \mathbf{p}(0),\mathbf{q}(0)\right|,\label{eq:HK_propagator}
\end{align}
where $(\mathbf{p}(t),\mathbf{q}(t))$ is the $2F$-dimensional classical
trajectory evolving from initial conditions $(\mathbf{p}(0),\mathbf{q}(0))$,
and $S_{t}$ is the corresponding classical action. \equationabb
(\ref{eq:HK_propagator}) also contains the HK prefactor,
\begin{align}
C_{t}(\mathbf{p}(0),\mathbf{q}(0))=\sqrt{\frac{1}{2^{F}}\text{det}\left[\frac{\partial\mathbf{q}(t)}{\partial\mathbf{q}(0)}+\frac{\partial\mathbf{p}(t)}{\partial\mathbf{p}(0)}-\text{i}\hbar\boldsymbol{\gamma}\frac{\partial\mathbf{q}(t)}{\partial\mathbf{p}(0)}+\frac{\text{i}}{\hbar\boldsymbol{\gamma}}\frac{\partial\mathbf{p}(t)}{\partial\mathbf{q}(0)}\right]}\label{eq:HK_prefactor}
\end{align}
which accounts for second-order quantum delocalizations around the
classical paths. Finally, the coherent state basis set in position
representation for many degrees of freedom is given by the direct
product of one-dimensional coherent states, 
\begin{align}
\left\langle \mathbf{x}|\mathbf{p},\mathbf{q}\right\rangle =\left(\frac{\det(\boldsymbol{\gamma})}{\pi^{F}}\right)^{1/4}\exp\left[-\frac{1}{2}\left(\mathbf{x}-\mathbf{q}\right){}^{\text{T}}\boldsymbol{\gamma}\left(\mathbf{x}-\mathbf{q}\right)+\frac{\text{i}}{\hbar}\mathbf{p}^{\text{T}}\left(\mathbf{x}-\mathbf{q}\right)\right]\label{eq:HK_gaussians}
\end{align}
where $\boldsymbol{\gamma}$ is a diagonal matrix containing $F$
time independent width parameters.

While the semiclassical approximation of the propagator in \equationabb
(\ref{eq:HK_propagator}) in principle allows for the inclusion of
an arbitrary number of DOFs, practical applications are limited by
the need to converge the phase space integral. Therefore, we will
now present two methods that are aimed at accelerating the numerical
Monte Carlo phase space integration of \equationabb (\ref{eq:HK_propagator}).
The first step is the introduction of a time averaging integral \cite{Elran1999,Kaledin2003},
which is applied to \equationabb (\ref{eq:IE_dyn}) and yields a
semiclassical approximation with a pre-averaged phase space integrand,
\begin{align}
\begin{split}I(E)= & \frac{1}{\left(2\pi\hbar\right)^{F}}\int\text{d}\mathbf{p}(0)\int\text{d}\mathbf{q}(0)\frac{1}{\pi\hbar T}\ \text{Re}\int\limits _{0}^{T}\text{d}t_{1}\int\limits _{t_{1}}^{T}\text{d}t_{2}\ C_{t_{2}}(\mathbf{p}(t_{1}),\mathbf{q}(t_{1}))\\
 & \times\left\langle \chi\middle|\mathbf{p}(t_{2}),\mathbf{q}(t_{2})\right\rangle \text{e}^{\text{i}\left[S_{t_{2}}\left(\mathbf{p}(0),\mathbf{q}(0)\right)+Et_{2}\right]/\hbar}\left[\left\langle \chi\middle|\mathbf{p}(t_{1}),\mathbf{q}(t_{1})\right\rangle \text{e}^{\text{i}\left[S_{t_{1}}\left(\mathbf{p}(0),\mathbf{q}(0)\right)+Et_{1}\right]/\hbar}\right]^{*}
\end{split}
\label{eq:HK_nonsep}
\end{align}
In order to recover a single time integration as in \equationabb
(\ref{eq:IE_dyn}), Kaledin and Miller have suggested the so-called
separable approximation \cite{Kaledin2003}, where the prefactor is
written as $C_{t_{2}}(\mathbf{p}(t_{1}),\mathbf{q}(t_{1}))\approx\exp\left[\text{i}(\phi(t_{2})-\phi(t_{1}))/\hbar\right]$,
and $\phi(t)/\hbar=\text{phase}\left[C_{t}(\mathbf{p}(0),\mathbf{q}(0))\right]$.
This procedure is exact in the harmonic limit and results in the expression
\begin{align}
 & I(E)=\frac{1}{\left(2\pi\hbar\right)^{F}}\frac{1}{2\pi\hbar T}\int\text{d}\mathbf{p}(0)\int\text{d}\mathbf{q}(0)\left|\int\limits _{0}^{T}\text{d}t\left\langle \chi\middle|\mathbf{p}(t),\mathbf{q}(t)\right\rangle \text{e}^{\text{i}\left[S_{t}\left(\mathbf{p}(0),\mathbf{q}(0)\right)+Et+\phi_{t}\left(\mathbf{p}(0),\mathbf{q}(0)\right)\right]/\hbar}\right|^{2}\label{eq:HK_sep}
\end{align}
which contains only a single and positive-definite phase space integrand
that is expected to be more stable numerically than the two-time integration
in \equationabb (\ref{eq:HK_nonsep}). While clearly less computationally
demanding than \equationabb (\ref{eq:HK_nonsep}), the separable
approximation in \equationabb (\ref{eq:HK_sep}) has also turned
out to be very accurate for a number of molecular dynamics applications.
\cite{Kaledin2003,Kaledin2003-1,Ceotto2011,Gabas2017,DiLiberto2016,Buchholz2016,Tamascelli2014,Conte2013,Ceotto2013,Zhuang2013,Ceotto2011-1,Ceotto2010,Ceotto2009,Ceotto2009-1}

The second step towards making the dynamics of larger systems accessible
is to invoke the mixed approximation. To this end, we use the semiclassical
hybrid dynamics idea to divide the $2F$ phase space variables into
$2F_{\text{hk}}$ for the system space and $2F_{\text{tg}}$ for the
bath phase space. Only the system part, denoted by the subscript hk,
is then treated on the HK level of accuracy, whereas the simpler single-trajectory
TGWD approximation is used for the bath DOFs, which are denoted by
the subscript tg. This separation is made only for the semiclassical
expression, while the underlying classical dynamics is not modified.
We now assume a reference state of Gaussian form, $\left|\chi\right\rangle =\left|\mathbf{p}_{\text{eq}},\mathbf{q}_{\text{eq}}\right\rangle $,
where $\mathbf{q_{\text{eq}}}$ is the equilibrium position and $\mathbf{p}_{\text{eq}}$
is the momentum corresponding to some eigenenergy. In the mixed approximation,
the initial phase space coordinates are
\begin{align}
\mathbf{p}_{\text{eq}}(0)=\left(\begin{matrix}\mathbf{p}_{\text{hk}}(0)\\
\mathbf{p}_{\text{eq},\:\text{tg}}(0)
\end{matrix}\right),\qquad\mathbf{q}_{\text{eq}}(0)=\left(\begin{matrix}\mathbf{q}_{\text{hk}}(0)\\
\mathbf{q}_{\text{eq},\:\text{tg}}(0)
\end{matrix}\right).\label{eq:HYB_initial}
\end{align}
 Only the HK initial conditions $\left(\mathbf{p}_{\text{hk}}(0),\mathbf{q}_{\text{hk}}(0)\right)$
are found by Monte Carlo sampling around $\left(\mathbf{p}_{\text{eq},\:\text{hk}},\mathbf{q}_{\text{eq},\:\text{hk}}\right)$,
while the bath starting coordinates are always at the equilibrium
positions, $\left(\mathbf{p}_{\text{eq},\:\text{tg}}(0),\mathbf{q}_{\text{eq},\:\text{tg}}(0)\right)=\left(\mathbf{p}_{\text{eq},\:\text{tg}},\mathbf{q}_{\text{eq},\:\text{tg}}\right)$.
Since the TGWD is exact for harmonic potentials, this division should
accurately reproduce the contributions of weakly coupled bath DOFs
close to their potential minimum. With this separation in place, we
expand the classical trajectories and the action to first and second
order, respectively, in the displacement coordinates of the bath subspace
\begin{align}
\delta\mathbf{p}_{\text{tg}}=\mathbf{p}_{\text{tg}}(0)-\mathbf{p}_{\text{eq},\:\text{tg}}(0),\qquad\delta\mathbf{q}_{\text{tg}}=\mathbf{q}_{\text{tg}}(0)-\mathbf{q}_{\text{eq},\:\text{tg}}(0).\label{eq:HYB_displacement}
\end{align}
This approximates the exponent in \equationabb (\ref{eq:HK_sep})
such that the phase space integration over the original bath initial
conditions $(\mathbf{p}_{\text{tg}}(0),\mathbf{q}_{\text{tg}}(0))$
can be performed analytically as a Gaussian integral. The expanded
classical trajectories become 
\begin{align}
\begin{split}\mathbf{p}(t)= & \left(\begin{matrix}\mathbf{p}_{\text{hk}}(t)\\
\mathbf{p}_{\text{tg}}(t)
\end{matrix}\right)=\mathbf{p}_{\text{eq}}(t)+\mathbf{m}_{11}(t)\delta\mathbf{p}_{\text{tg}}+\mathbf{m}_{12}(t)\delta\mathbf{q}_{\text{tg}}\\
\mathbf{q}(t)= & \left(\begin{matrix}\mathbf{q}_{\text{hk}}(t)\\
\mathbf{q}_{\text{tg}}(t)
\end{matrix}\right)=\mathbf{q}_{\text{eq}}(t)+\mathbf{m}_{21}(t)\delta\mathbf{p}_{\text{tg}}+\mathbf{m}_{22}(t)\delta\mathbf{q}_{\text{tg}},
\end{split}
\label{eq:HYB_trajectories}
\end{align}
and the action is
\begin{align}
\begin{split}S_{t}\left(\mathbf{p}\left(0\right),\mathbf{q}\left(0\right)\right) & =S_{t}\left(\mathbf{p}_{\text{hk}}\left(0\right),\mathbf{q}_{\text{hk}}\left(0\right),\mathbf{p}_{\text{eq},\:\text{tg}}\left(0\right),\mathbf{q}_{\text{eq},\:\text{tg}}\left(0\right)\right)\\
 & +\mathbf{p}_{\text{eq}}^{\text{T}}(t)\mathbf{m}_{21}\left(t\right)\delta\mathbf{p}_{\text{tg}}+\left(\mathbf{p}_{\text{eq}}^{\text{T}}(t)\mathbf{m}_{22}\left(t\right)-\mathbf{p}_{\text{eq},\:0,\:\text{tg}}^{\text{T}}\right)\delta\mathbf{q}_{\text{tg}}\\
 & +\frac{1}{2}\delta\mathbf{p}_{\text{tg}}^{\text{T}}\mathbf{m}_{11}^{\text{T}}\left(t\right)\mathbf{m}_{21}\left(t\right)\delta\mathbf{p}_{\text{tg}}+\frac{1}{2}\delta\mathbf{q}_{\text{tg}}^{\text{T}}\mathbf{m}_{12}^{\text{T}}\left(t\right)\mathbf{m}_{22}\left(t\right)\delta\mathbf{q}_{\text{tg}}\\
 & +\delta\mathbf{q}_{\text{tg}}^{\text{T}}\mathbf{m}_{12}^{\text{T}}\left(t\right)\mathbf{m}_{21}\left(t\right)\delta\mathbf{p}_{\text{tg}}
\end{split}
\label{eq:HYB_action}
\end{align}
The $\mathbf{m}_{ij}$ in \equationsabb (\ref{eq:HYB_trajectories})
and (\ref{eq:HYB_action}) are non-square $F\times F_{\text{tg}}$
submatrices of the stability matrix, 
\begin{align}
\begin{split}\mathbf{m}_{11}(t)= & \frac{\partial\mathbf{p}_{\text{eq}}(t)}{\partial\mathbf{p}_{\text{eq},\:\text{tg}}(0)},\qquad\mathbf{m}_{12}(t)=\frac{\partial\mathbf{p}_{\text{eq}}(t)}{\partial\mathbf{q}_{\text{eq},\:\text{tg}}(0)},\\
\mathbf{m}_{21}(t)= & \frac{\partial\mathbf{q}_{\text{eq}}(t)}{\partial\mathbf{p}_{\text{eq},\:\text{tg}}(0)},\qquad\mathbf{m}_{22}(t)=\frac{\partial\mathbf{q}_{\text{eq}}(t)}{\partial\mathbf{q}_{\text{eq},\:\text{tg}}(0)}.
\end{split}
\label{eq:HYB_mij}
\end{align}
They will be used only for the TG part of the mixed TA-SCIVR integrand,
while the phase $\phi_{t}$ of the HK prefactor still comprises the
full $F\times F$ matrices from \equationabb (\ref{eq:HK_prefactor}).
After unraveling the modulus in \equationabb (\ref{eq:HK_sep}),
inserting \equationsabb (\ref{eq:HYB_trajectories}) and (\ref{eq:HYB_action}),
the phase space integration over the TG DOFs can be performed analytically
as a Gaussian integral. This results in the mixed TA-SCIVR expression
\begin{align}
I(E)= & \frac{1}{\left(2\hbar\right)^{F}}\frac{1}{\pi^{F_{\text{hk}}}}\frac{\text{Re}}{\pi\hbar T}\int\text{d}\mathbf{p}_{\text{hk}}\left(0\right)\int\text{d}\mathbf{q}_{\text{hk}}\left(0\right)\int_{0}^{T}\text{d}t_{1}\int_{t_{1}}^{T}\text{d}t_{2}\nonumber \\
 & \times\text{e}^{\text{i}\left[E\left(t_{1}-t_{2}\right)+\phi_{t_{1}}\left(\mathbf{p}_{\text{eq}}\left(0\right),\:\mathbf{q}_{\text{eq}}\left(0\right)\right)-\phi_{t_{2}}\left(\mathbf{p}_{\text{eq}}\left(0\right),\:\mathbf{q}_{\text{eq}}\left(0\right)\right)+S_{t_{1}}\left(\mathbf{p}_{\text{eq}}\left(0\right),\:\mathbf{q}_{\text{eq}}\left(0\right)\right)-S_{t_{2}}\left(\mathbf{p}_{\text{eq}}\left(0\right),\:\mathbf{q}_{\text{eq}}\left(0\right)\right)\right]/\hbar}\nonumber \\
 & \times\left\langle \mathbf{p}_{\text{eq},\:\text{hk}},\mathbf{q}_{\text{eq},\:\text{hk}}\middle|\mathbf{p}_{\text{eq},\:\text{hk}}\left(t_{1}\right),\mathbf{q}_{\text{eq},\:\text{hk}}\left(t_{1}\right)\right\rangle \left\langle \mathbf{p}_{\text{eq},\:\text{hk}}\left(t_{2}\right),\mathbf{q}_{\text{eq},\:\text{hk}}\left(t_{2}\right)\middle|\mathbf{p}_{\text{eq},\:\text{hk}},\mathbf{q}_{\text{eq},\:\text{hk}}\right\rangle \label{eq:HYB_nonsep}\\
 & \times\sqrt{\frac{1}{\det\left(\mathbf{A}\left(t_{1}\right)+\mathbf{A}^{*}\left(t_{2}\right)\right)}}\nonumber \\
 & \times\left\langle \mathbf{p}_{\text{eq},\:\text{tg}},\mathbf{q}_{\text{eq},\:\text{tg}}\middle|\mathbf{p}_{\text{eq},\:\text{tg}}\left(t_{1}\right),\mathbf{q}_{\text{eq},\:\text{tg}}\left(t_{1}\right)\right\rangle \left\langle \mathbf{p}_{\text{eq},\:\text{tg}}\left(t_{2}\right),\mathbf{q}_{\text{eq},\:\text{tg}}\left(t_{2}\right)\middle|\mathbf{p}_{\text{eq},\:\text{tg}},\mathbf{q}_{\text{eq},\:\text{tg}}\right\rangle \\
 & \times\mbox{exp}\left\{ \frac{1}{4}\left(\mathbf{b}_{t_{1}}+\mathbf{b}_{t_{2}}^{*}\right)^{\text{T}}\left(\mathbf{A}\left(t_{1}\right)+\mathbf{A}^{*}\left(t_{2}\right)\right)^{-1}\left(\mathbf{b}_{t_{1}}+\mathbf{b}_{t_{2}}^{*}\right)\right\} ,\nonumber 
\end{align}
 which contains some newly defined expressions, namely, the symmetric
$2F_{\text{tg}}\times2F_{\text{tg}}$ matrix $\mathbf{A}(t)$ with
blocks
\begin{align}
\begin{split}\mathbf{A}_{11}(t)= & \frac{1}{4}\mathbf{m}_{21}^{\text{T}}\left(t\right)\mathbf{\boldsymbol{\gamma}}\mathbf{m}_{21}\left(t\right)+\frac{1}{4\hbar^{2}}\mathbf{m}_{11}^{\text{T}}\left(t\right)\mathbf{\boldsymbol{\gamma}}^{-1}\mathbf{m}_{11}\left(t\right)\\
\mathbf{A}_{12}(t)= & \frac{1}{4}\mathbf{m}_{21}^{\text{T}}\left(t\right)\boldsymbol{\gamma}\mathbf{m}_{22}\left(t\right)+\frac{1}{4\hbar^{2}}\mathbf{m}_{11}^{\text{T}}\left(t\right)\boldsymbol{\gamma}^{-1}\mathbf{m}_{12}\left(t\right)\\
\mathbf{A}_{21}(t)= & \frac{1}{4}\mathbf{m}_{22}^{\text{T}}\left(t\right)\boldsymbol{\gamma}\mathbf{m}_{21}\left(t\right)+\frac{1}{4\hbar^{2}}\mathbf{m}_{12}^{\text{T}}\left(t\right)\boldsymbol{\gamma}^{-1}\mathbf{m}_{11}\left(t\right)\\
\mathbf{A}_{22}(t)= & \frac{1}{4}\mathbf{m}_{22}^{\text{T}}\left(t\right)\boldsymbol{\gamma}\mathbf{m}_{22}\left(t\right)+\frac{1}{4\hbar^{2}}\mathbf{m}_{12}^{\text{T}}\left(t\right)\mathbf{\boldsymbol{\gamma}}^{-1}\mathbf{m}_{12}\left(t\right),
\end{split}
\label{eq:HYB_A_matrix}
\end{align}
and the $2F_{\text{tg}}$-dimensional vector $\mathbf{b}(t)\equiv\left(\mathbf{b}_{1,t}^{\text{T}},\mathbf{b}_{2,t}^{\text{T}}\right)^{\text{T}}$
with subvectors 
\begin{align}
\begin{split}\mathbf{b}_{1,t}^{\text{T}}= & -\frac{1}{2}\left(\mathbf{q}\left(t\right)-\mathbf{q}\left(0\right)\right)^{\text{T}}\left[\mathbf{\boldsymbol{\gamma}}\mathbf{m}_{21}\left(t\right)+\frac{\text{i}}{\hbar}\mathbf{m}_{11}\left(t\right)\right]\\
 & -\frac{1}{2\hbar^{2}}\left(\mathbf{p}\left(t\right)-\mathbf{p}\left(0\right)\right)^{\text{T}}\left[\mathbf{\boldsymbol{\gamma}}^{-1}\mathbf{m}_{11}\left(t\right)-\text{i}\hbar\mathbf{m}_{21}\left(t\right)\right]\\
\mathbf{b}_{2,t}^{\text{T}}= & -\frac{1}{2}\left(\mathbf{q}\left(t\right)-\mathbf{q}\left(0\right)\right)^{\text{T}}\left[\mathbf{\boldsymbol{\gamma}}\mathbf{m}_{22}\left(t\right)+\frac{\text{i}}{\hbar}\mathbf{m}_{12}\left(t\right)\right]\\
 & -\frac{1}{2\hbar^{2}}\left(\mathbf{p}\left(t\right)-\mathbf{p}\left(0\right)\right)^{\text{T}}\left[\mathbf{\boldsymbol{\gamma}}^{-1}\mathbf{m}_{12}\left(t\right)-\text{i}\hbar\mathbf{m}_{22}\left(t\right)\right].
\end{split}
\label{eq:HYB_b_vector}
\end{align}
The expressions for $\mathbf{A}(t)$ and $\mathbf{b}(t)$ differ from
our first publication on this matter\cite{Buchholz2016}, as we have
left out two constant imaginary contributions in \equationsabb (\ref{eq:HYB_A_matrix})
and (\ref{eq:HYB_b_vector}) that cancel out in the phase space integrand
in \equationabb (\ref{eq:HYB_nonsep}). Another difference to Ref$.\ $\onlinecite{Buchholz2016}
is that we have explicitly written out the scalar quantity $c_{t}$,
which has been defined there and which contained the action as well
as the overlap of the TG part with the initial state.

Comparing \equationabb (\ref{eq:HYB_nonsep}) to the full HK expressions
(\ref{eq:HK_nonsep}) and (\ref{eq:HK_sep}), we have achieved a reduction
in dimensionality of the phase space that has to be sampled over.
The loss in accuracy is expected to be minimal, as the bath DOFs that
are treated on the TG level are weakly coupled and therefore close
to harmonic behavior. Again, we stress that there is no decoupling
of the underlying classical dynamics.

While the reduced Monte Carlo sampling is clearly advantageous for
numerical efficiency, it has come at the price of reintroducing two
time integrations in \equationabb (\ref{eq:HYB_nonsep}). The integration
itself poses no difficulty, as it is simply a two-dimensional Fourier
transformation, but calculating the integrand for $N_{\text{steps}}^{2}$
time steps takes a lot of computational time. Therefore, it is highly
desirable to find an expression with only a single time integration.
In the spirit of the original separable approximation that lead from
\equationabb (\ref{eq:HK_nonsep}) to \equationabb (\ref{eq:HK_sep}),
we proceed by assuming a form for the TG exponent and the TG prefactor
that is exact for the harmonic oscillator\cite{Buchholz2016}. The
separable form of the exponent reads 
\begin{align}
\begin{split} & \frac{1}{4}\left(\mathbf{b}_{t_{1}}+\mathbf{b}_{t_{2}}^{*}\right)^{\text{T}}\left(\mathbf{A}\left(t_{1}\right)+\mathbf{A}^{*}\left(t_{2}\right)\right)^{-1}\left(\mathbf{b}_{t_{1}}+\mathbf{b}_{t_{2}}^{*}\right)\\
 & \qquad\approx\frac{1}{4}\mathbf{b}_{t_{1}}^{\text{T}}\left(\mathbf{A}\left(t_{1}\right)+\mathbf{A}^{*}\left(t_{1}\right)\right)^{-1}\mathbf{b}_{t_{1}}+\frac{1}{4}\left[\mathbf{b}_{t_{2}}^{\text{T}}\left(\mathbf{A}\left(t_{2}\right)+\mathbf{A}^{*}\left(t_{2}\right)\right)^{-1}\mathbf{b}_{t_{2}}\right]^{*},
\end{split}
\label{eq:HYB_sep_exponent}
\end{align}
and the TG prefactor is separated in the fashion of a geometric average,
\begin{align}
\frac{1}{\sqrt{\det\left(\mathbf{A}\left(t_{1}\right)+\mathbf{A}^{*}\left(t_{2}\right)\right)}}\approx\left(\frac{1}{\det\left(\mathbf{A}\left(t_{1}\right)+\mathbf{A}\left(t_{1}\right)\right)}\right)^{1/4}\left(\frac{1}{\det\left(\mathbf{A}\left(t_{2}\right)+\mathbf{A}\left(t_{2}\right)\right)}\right)^{1/4}.\label{eq:HYB_sep_prefactor}
\end{align}
With that, we arrive at the desired separable mixed TA-SCIVR, 
\begin{align}
I(E)= & \ \frac{1}{\left(2\hbar\right)^{F}}\frac{1}{\pi^{F_{\text{hk}}}}\frac{1}{2\pi\hbar T}\int\text{d}\mathbf{p}_{\text{hk}}\left(0\right)\int\text{d}\mathbf{q}_{\text{hk}}\left(0\right)\left|\int_{0}^{T}\text{d}t\:\text{e}^{\text{i}\left[Et+\phi_{t}\left(\mathbf{p}_{\text{eq}}\left(0\right),\:\mathbf{q}_{\text{eq}}\left(0\right)\right)+S_{t}\left(\mathbf{p}_{\text{eq}}\left(0\right),\:\mathbf{q}_{\text{eq}}\left(0\right)\right)\right]/\hbar}\right.\nonumber \\
 & \times\left\langle \mathbf{p}_{\text{eq},\:\text{hk}},\mathbf{q}_{\text{eq},\:\text{hk}}\middle|\mathbf{p}_{\text{eq},\:\text{hk}}\left(t\right),\mathbf{q}_{\text{eq},\:\text{hk}}\left(t\right)\right\rangle \left\langle \mathbf{p}_{\text{eq},\:\text{tg}},\mathbf{q}_{\text{eq},\:\text{tg}}\middle|\mathbf{p}_{\text{eq},\:\text{tg}}\left(t\right),\mathbf{q}_{\text{eq},\:\text{tg}}\left(t\right)\right\rangle \label{eq:HYB_sep}\\
 & \times\left.\frac{1}{\left[\det\left(\mathbf{A}\left(t\right)+\mathbf{A}^{*}\left(t\right)\right)\right]^{1/4}}\exp\left\{ \frac{1}{4}\mathbf{b}_{t}^{\text{T}}\left(\mathbf{A}\left(t\right)+\mathbf{A}^{*}\left(t\right)\right)^{-1}\mathbf{b}_{t}\right\} \right|^{2}\nonumber 
\end{align}
As we have seen for two-dimensional and three-dimensional model systems\cite{Buchholz2016},
this approximation reproduces both system and bath peaks precisely
when compared with exact quantum dynamics results, and reaches tight
convergence within a considerably shorter amount of time than the
separable TA-SCIVR from \equationabb (\ref{eq:HK_sep}).

In the next section, we show that this high accuracy is also achieved
for larger systems, and we then go on to use the mixed approximation
to investigate the influence of the Caldeira-Leggett counter term
on the frequency shift of an anharmonic Morse system. We note in passing
that a linearization along the lines of linearized SC-IVRs (LSC-IVR)
is not possible, because we just have a single time-evolution operator
in our starting expression (\ref{eq:IE_def}), while the LSC-IVR is
propagating densities. \cite{Bonnet2013,Liu2015,Liu2007-1,Liu2011} 

\section{Model: Morse Oscillator Coupled to a Caldeira-Leggett Bath}

\label{sec:model}

In order to test the accuracy of the mixed TA-SCIVR, we use a Morse
oscillator coupled bilinearly to a Caldeira-Leggett (CL) bath of harmonic
oscillators. The Hamiltonian in atomic units has the form 
\begin{align}
H=\frac{p_{\text{s}}^{2}}{2m_{\text{s}}}+V_{\text{s}}(s)+\sum\limits _{i=1}^{F_{\text{b}}}\left[\frac{p_{i}^{2}}{2}+\frac{1}{2}\omega_{i}^{2}y_{i}^{2}+c_{i}y_{i}\left(s-s_{\text{eq}}\right)+\frac{1}{2}\frac{c_{i}^{2}}{\omega_{i}^{2}}\left(s-s_{\text{eq}}\right)^{2}\right]\label{eq:CL_Hamiltonian}
\end{align}
where the Morse potential for the system coordinate $s$ is 
\begin{align}
V_{\text{s}}(s)=D_{\text{e}}\left(1-\text{e}^{-\alpha\left(s-s_{\text{eq}}\right)}\right)^{2},\label{eq:CL_Morse_pot}
\end{align}
and we take the parameters of molecular iodine\cite{Ovchinnikov1996}
for the dissociation energy $D_{\text{e}}=0.057\ \text{a.u.}$, for
the equilibrium distance $s_{\text{eq}}=5.001\ \text{a.u.}$, and
for the range parameter $\alpha=0.983\ \text{a.u.}$ The reduced mass
of the Morse oscillator is $m_{\text{s}}=m_{\text{r}}=1.165\times10^{5}\ \text{a.u.}$
There is an analytic solution for the eigenenergies of the Morse potential
which we will need later for comparison,
\begin{align}
E_{n}=\omega_{\text{e}}\left(n+\frac{1}{2}\right)-\omega_{\text{e}}x_{\text{e}}\left(n+\frac{1}{2}\right)^{2},\label{eq:CL_Morse_pot_solution}
\end{align}
where $\omega_{\text{e}}=\alpha/\sqrt{m_{\text{r}}/(2D_{\text{e}})}$
is the frequency of the harmonic approximation to the Morse potential,
and $x_{\text{e}}=\omega_{\text{e}}/(4D_{\text{e}})$ is the anharmonicity
parameter. For the iodine parameters above, these quantities become
$\omega_{\text{e}}=9.724\times10^{-4}\ \text{a.u.}$ and $x_{\text{e}}=4.264\times10^{-3}\ \text{a.u.}$

The bath part of the Hamiltonian in \equationabb (\ref{eq:CL_Hamiltonian})
consists of the bath kinetic energy, the bilinear system-bath coupling,
where $y_{i}$ denotes the bath DOFs, and the Caldeira-Leggett counter
term. The latter is introduced in order to prevent a renormalization
of the potential \cite{Ingold2002}, and we will look into its influence
on the system spectrum in the next section. Following Refs.\ \onlinecite{Wang2001}
and \onlinecite{Goletz2009}, we use an Ohmic spectral density with
an exponential cutoff, 
\begin{align}
J_{\text{e}}(\omega)=\eta\omega\text{e}^{-\omega/\omega_{\text{c}}},\label{eq:CL_bath_specdensity_discrete}
\end{align}
with the system-bath coupling strength $\eta$ and a cutoff frequency
$\omega_{\text{c}}$. In discretized form, the density is defined
as 
\begin{align}
J(\omega)=\frac{\pi}{2}\sum\limits _{i=1}^{F_{\text{b}}}\frac{c_{i}^{2}}{\omega_{i}}\delta(\omega-\omega_{i}),\label{eq:CL_bath_specdensity_continuous}
\end{align}
and the coupling coefficients $c_{i}$ in \equationabb (\ref{eq:CL_Hamiltonian})
are chosen such that it becomes equivalent to the continuous form
in the limit of infinitely many bath oscillators, 
\begin{align}
c_{i}^{2}=\frac{2}{\pi}\omega_{i}\frac{J_{\text{e}}(\omega_{i})}{\rho(\omega_{i})},
\end{align}
with the frequency density defined by the condition 
\begin{align}
\int\limits _{0}^{\omega_{i}}\text{d}\omega\ \rho(\omega)=i\qquad\text{for }i=1,\dots,F_{\text{b}}.\label{eq:CL_bath_freqdensity_def}
\end{align}
Here, we choose it as 
\begin{align}
\rho(\omega)=a\frac{J_{\text{e}}(\omega)}{\omega}\label{eq:CL_bath_freqdensity_choice}
\end{align}
where $a$ is a normalization coefficient to ensure that $i=F_{\text{b}}$
if the largest bath frequency $\omega_{i}=\omega_{\text{max}}$ is
chosen in \equationabb (\ref{eq:CL_bath_freqdensity_def}), and it
amounts to 
\begin{align}
a=\frac{F_{\text{b}}}{\eta\omega_{\text{c}}}\frac{1}{1-\text{e}^{-\omega_{\text{max}}/\omega_{\text{c}}}}.
\end{align}
With \equationsabb (\ref{eq:CL_bath_specdensity_continuous}), (\ref{eq:CL_bath_freqdensity_def})
and (\ref{eq:CL_bath_freqdensity_choice}), one finds the discrete
frequencies as 
\begin{align}
\omega_{i}=-\omega_{\text{c}}\ln\left(1-\frac{i(1-\text{e}^{-\omega_{\text{max}}/\omega_{\text{c}}})}{F_{\text{b}}}\right).
\end{align}
If both the cutoff and the maximum frequency of the bath are chosen
much smaller than the system frequency, about $F_{\text{b}}=20$ bath
oscillators have been shown to be sufficient to reproduce a continuous
bath\cite{Wang2001}. The semiclassical hybrid approach in particular
has already turned out to provide an adequate description for the
short-time decay of quantum coherence of this specific system-bath
problem\cite{Goletz2009}. We also choose this frequency density because
it allows to set up a bath containing not only many low-frequency
modes, but also a few oscillators with frequencies close to the system
frequency. A thorough study comparing different spectral densities
with their advantages and drawbacks is given in Ref$.\ $\onlinecite{Goletz2010}.

\section{Results and Discussion}

\label{sec:results}

The main objective of this paper is the description of the frequency
shift of a Morse oscillator coupled to a CL bath using the mixed TA-SCIVR.
In order to demonstrate the very good accuracy of the mixed approach,
and in particular the separable approximation, we first discuss results
for a ten-dimensional bath where TA HK results according to \equationabb
(\ref{eq:HK_sep}) can still be found relatively easily. After making
this comparison, we will turn to baths with up to 60 DOFs and different
bath parameters to show their influence on the system spectrum. A
specific focus will be on the role of the CL counter term for the
anharmonic spectrum.

We employ two different frequency combinations, one with a resonant
maximum frequency, $\omega_{\text{c}}=0.5\ \omega_{\text{s}}$ and
$\omega_{\max}=\omega_{\text{s}}$, and a low-frequency bath with
$\omega_{\text{c}}=0.1\ \omega_{\text{s}}$ and $\omega_{\max}=0.2\ \omega_{\text{s}}$.
We choose two different effective coupling parameters $\eta_{\text{eff}}=\eta/(m_{\text{s}}\omega_{\text{s}})$,
namely, $\eta_{\text{eff}}=0.5$ and $\eta_{\text{eff}}=2.0$ for
the bath with small cutoff frequency and $\eta_{\text{eff}}=0.1$
and $\eta_{\text{eff}}=0.5$ for the big cutoff. The system DOF initially
is at equilibrium with nonzero momentum, $(0,\sqrt{m_{\text{s}}\omega_{\text{s}}})$,
while the bath oscillators are located at $(0,0)$ because otherwise
the spectrum becomes very noisy due to the huge number of excited
bath peaks (for an example with 19 DOFs, see \figureabb 3 of Ref.\ \onlinecite{Ceotto2011}).
In general, these simplified initial conditions might not be adequate
to describe the system frequency shift because possible anharmonic
contributions of the bath DOFs are neglected by a dynamics that explores
mainly the harmonic neighborhood of the potential minimum. For the
bilinearly coupled, harmonic CL bath, however, the difference in the
system frequencies arising from initially excited bath DOFs is negligible.
The number of semiclassical time steps is $N_{\text{steps}}=2^{14}$
and their length is $\Delta t=(2\pi/\omega_{\text{e}})/20$, resulting
in a frequency resolution of $1.2\times10^{-6}\ \text{a.u.}$ ($0.55\ \text{cm}^{-1}$).

\subsection{Morse Oscillator Coupled to Ten Harmonic Oscillators}

\begin{figure}[t!]
\centering \includegraphics[width=0.5\textwidth]{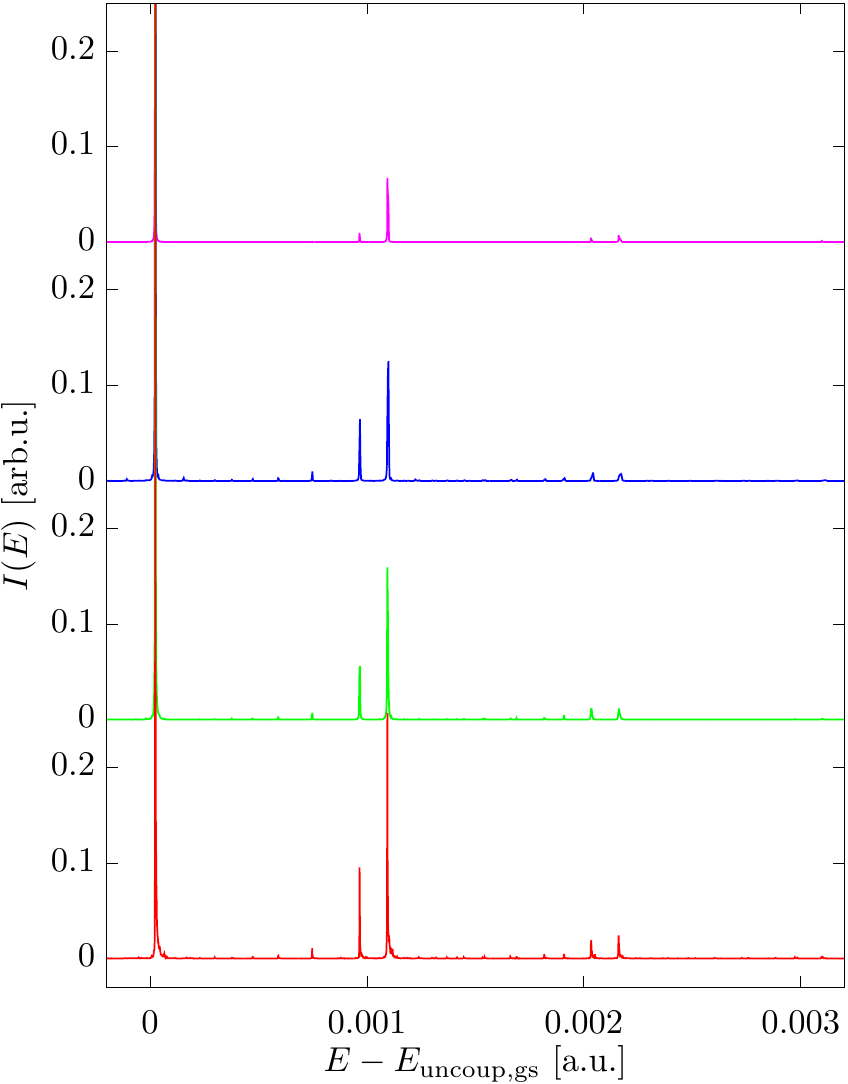}
\caption{\label{fig:morsecl_11D} Spectrum of an iodine-like Morse oscillator
coupled to a CL bath comprising 10 DOFs with $\omega_{\max}=\omega_{\text{s}}$,
$\omega_{\text{c}}=0.5\ \omega_{\text{s}}$ and coupling $\eta_{\text{eff}}=0.5$.
From top to bottom: separable TA mixed with 1 HK DOF (magenta) (\equationabb
(\ref{eq:HYB_sep})), full TA mixed with 1 HK DOF (blue) and full
TA mixed with 2 HK DOFs (green) (\equationabb (\ref{eq:HYB_nonsep})),
and the separable TA HK (red) using \equationabb (\ref{eq:HK_nonsep}).
All spectra are renormalized such that the groundstate peak has height
one. }
\end{figure}

\begin{figure}[t!]
\centering \includegraphics[width=0.5\textwidth]{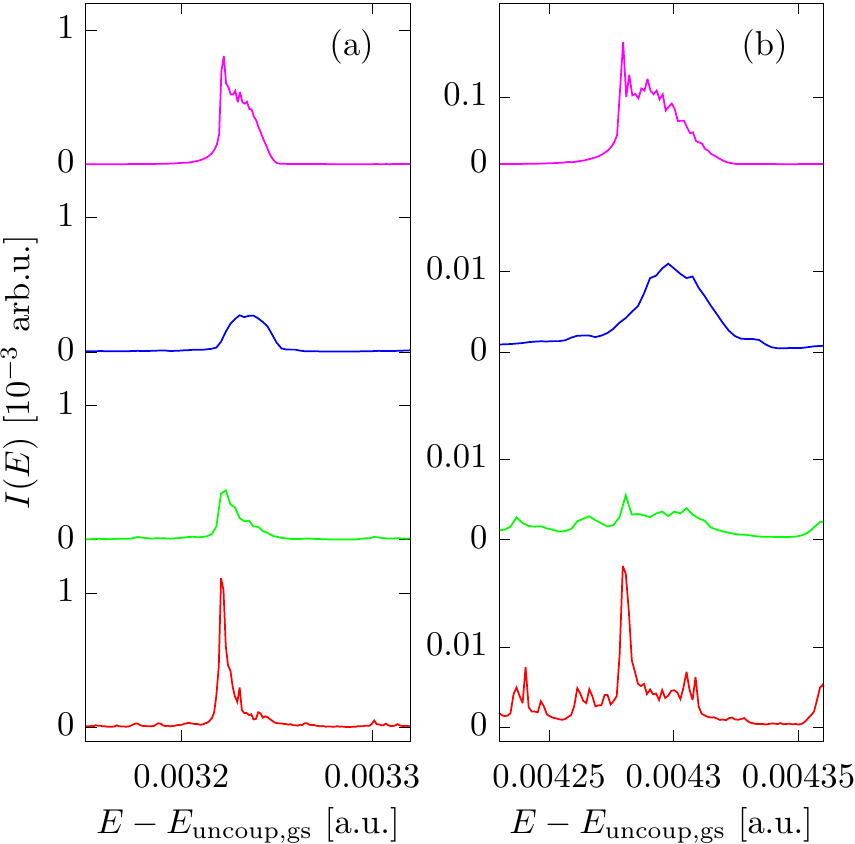}
\caption{\label{fig:morsecl_11D_details} The third (left) and fourth (right)
excitation of a Morse oscillator coupled to a CL bath. All bath parameters
and plot specifications are identical to \figureabb \ref{fig:morsecl_11D}.}
\end{figure}

\begin{figure}[t!]
\centering \includegraphics[width=0.5\textwidth]{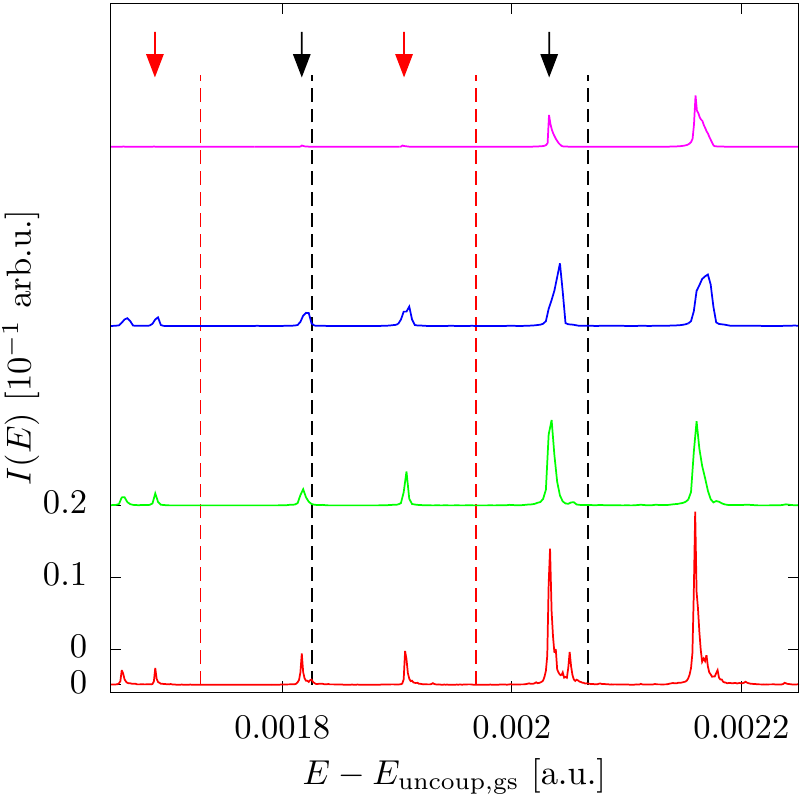}
\caption{\label{fig:morsecl_11D_bath} Bath excitations in the spectrum of
a Morse oscillator coupled to a CL bath. The respective rightmost
peak is the second excitation of the system, the remaining peaks originate
from bath modes. All bath parameters and plot specifications are identical
to \figureabb \ref{fig:morsecl_11D}. The arrows in panel designate
bath excitations of first order (fundamental) in black color and higher
order (overtones) in red color, dashed vertical lines show the respective
uncoupled counterpart.}
\end{figure}

We first discuss an interesting but relatively simple example, the
bath comprising ten oscillators with $\omega_{\text{c}}=0.5\ \omega_{\text{s}}$,
$\omega_{\max}=\omega_{\text{s}}$, and $\eta_{\text{eff}}=0.5$ (\figuresabb
\ref{fig:morsecl_11D} to \ref{fig:morsecl_11D_bath}). In \figureabb
\ref{fig:morsecl_11D}, we give an overview of results obtained with
the different methods. The degree of approximation always decreases
from top to bottom: the separable mixed approximation according to
\equationabb (\ref{eq:HYB_sep}) is indicated with magenta lines,
the full mixed approximation \equationabb (\ref{eq:HYB_nonsep})
is blue for one and green for two HK DOFs, and the reference separable
TA-SCIVR (\equationabb (\ref{eq:HK_nonsep}) is red. In the full
mixed approximation calculations, either only the Morse DOF is treated
with HK, or both Morse oscillator and the resonant bath mode, which
is expected to experience the strongest anharmonic driving by the
system. Only $10^{4}$ trajectories have been used in each case, both
to achieve reasonable computational costs and to work out the efficiency
of the new methods. In all spectral plots, we subtract the sum of
the ground state energies of the individual DOFs, 
\begin{align}
E_{\text{plot}}=E-E_{0}-\sum\limits _{i=1}^{F_{\text{b}}}\frac{\omega_{i}}{2},\label{eq:E_plot}
\end{align}
in order to make the net effect of the system-bath coupling visible
and to facilitate the comparison between baths with different parameters.
In Eq$.\ $(\ref{eq:E_plot}), $E_{0}$ is the ground state energy
of the Morse oscillator. 

Overall, agreement is very good between all methods. Bath peaks are
generally not very prominent because there is no initial excitation
in the bath; the only dynamics is induced by the system. This is reflected
especially in the reference TA-SCIVR and in the full mixed spectra
by the fact that the biggest bath peaks are those that correspond
to the modes whose frequency is closest to the system. By contrast,
just one bath peak from the resonant HO is featured significantly
in the separable mixed spectrum. 

\figureabb \ref{fig:morsecl_11D_details} highlights some details
of the spectra, namely, a zoom into the region of the third and fourth
excited system peak. The peaks in these pictures are three to five
orders of magnitude smaller than the groundstate and therefore quite
noisy, but they can nevertheless be identified as peaks. The full
mixed result with one HK DOF clearly disagrees with the reference
spectrum. However, this deviation can be removed by treating also
the most strongly coupled bath DOF on the HK level of accuracy, which
reproduces non-Gaussian distortions of the resonant bath mode. Another
way to include these distortions with just one HK DOF might have been
to use significantly more trajectories, thus sampling the resonant
bath mode indirectly by its coupling to the system mode via the classical
dynamics, as discussed in Ref. \onlinecite{Buchholz2012}. The separable
mixed results for those two peaks agree within the frequency resolution
with the reference full HK spectrum. In addition, they are better
converged than all the other methods, as can be seen in particular
for the fourth excited peak in \figureabb \ref{fig:morsecl_11D_details}(b).

In \figureabb \ref{fig:morsecl_11D_bath}, we put the bath excitations
under the spotlight. The rightmost peak is the second excited state
of the system, all remaining peaks are bath excitations (indicated
by red and black arrows). The dashed black lines have been obtained
by adding the bath frequencies $\omega_{i}$ to the first excited
system peak and thus illustrate where these bath peaks would be situated
if the bath frequencies remained unchanged by the dynamics. In the
same way, the dashed red lines show the expected uncoupled position
of higher order bath peaks. The rightmost red line, for example, shows
the second excited state of the HO with highest frequency. One sees
immediately that each bath peak lies to the left of the respective
dashed line, which means that all of these bath oscillators are redshifted.
Higher order bath excitations (red color arrows in \figureabb \ref{fig:morsecl_11D_bath})
are shifted further, as it is expected. As discussed in our first
paper on the mixed TA-SCIVR, the separable approximation that leads
to \equationabb (\ref{eq:HYB_sep}) entails a suppression of bath
excitations. Consequently, only the first excitation of the highest
frequency bath mode shows up significantly in the spectrum. Like the
excited states of the Morse oscillator, its position agrees closely
with the less approximate results. The other bath excitations are
strongly suppressed by the separable mixed method but they can still
be identified reliably upon closer inspection and turn out to be also
reproduced faithfully.

Due to its numerical advantages, we will perform exclusively separable
mixed calculations in the remainder of this paper, where we investigate
the system behavior for different bath characteristics.

\subsection{Frequency shifts for different bath sizes and role of the Caldeira-Leggett
counter term}

Having established that the separable mixed method offers the same
accuracy with respect to peak positions as the full HK treatment for
the CL system, we now increase the bath size up to 60 bath HOs. Again,
we use a very off resonant bath on the one hand and one with bath
frequencies up to the system frequency on the other. In addition,
we will analyse the influence of the CL counter term on the outcome
of the spectral calculations for our examples. We will undertake a
similar investigation as previously performed by Georgievskii and
Stuchebrukhov \cite{Georgievskii1990} and therefore look at the CL
model in the form of \equationabb (\ref{eq:CL_Hamiltonian}) as well
as a Hamiltonian without the counter term, 
\begin{align}
H=\frac{p_{\text{s}}^{2}}{2m_{\text{s}}}+V_{\text{s}}(s)
+\sum_{i=1}^{F_{\text{b}}}\left\{ \frac{p_{i}^{2}}{2}+\frac{1}{2}\omega_{i}^{2}y_{i}^{2}+c_{i}y_{i}\left(s-s_{\text{eq}}\right)\right\} .\label{eq:CL_Hamiltonian_noCT}
\end{align}

The following numerical investigations will comprise bath sizes of
10, 20, 40, and 60 DOFs, such that convergence with respect to the
number of bath HOs can be tested. As we are using the separable mixed
TA-SCIVR, we can keep the number of trajectories constant at $10^{4}$
for the differently sized baths. The number of HK DOFs has been either
one or two. Especially for the low frequency bath it was sufficient
to describe only the Morse oscillator with HK, while in the case of
the high bath cutoff it was helpful to include the resonant bath oscillator
into the HK part as well.

\begin{figure}[t!]
\centering \includegraphics[width=0.45\textwidth]{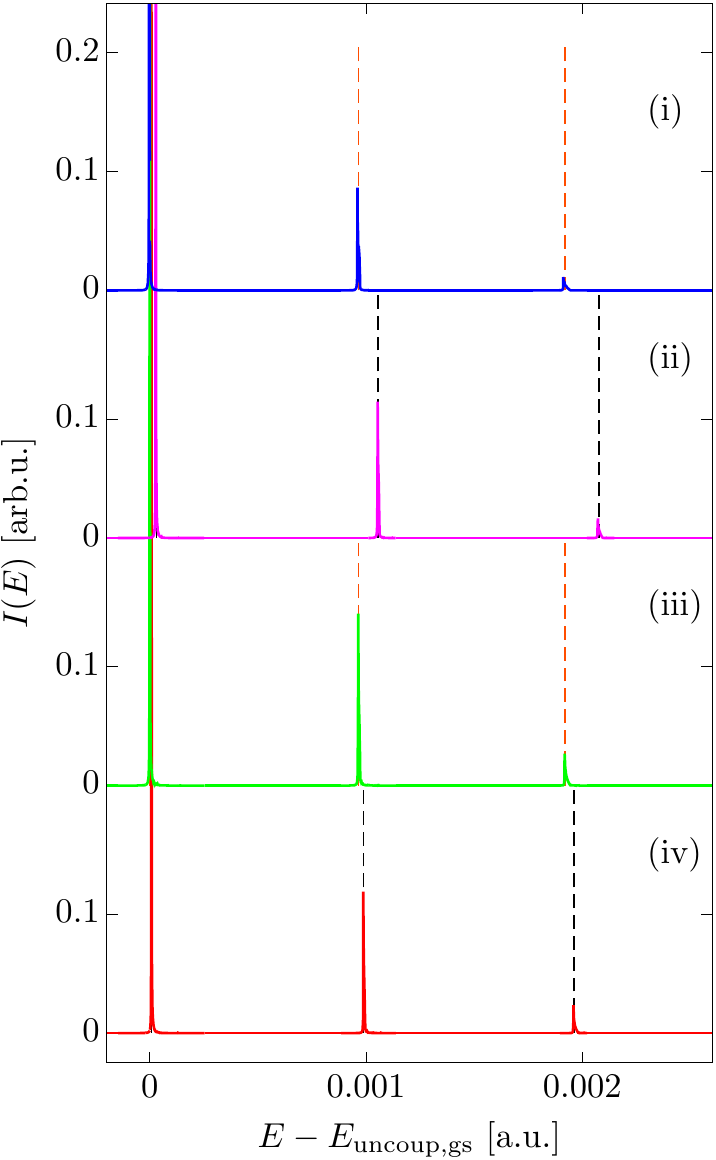}
\caption{Spectra for a Morse oscillator in a CL bath with 20 HOs for low cutoff
frequency. Bath parameters are $\omega_{\text{c}}=0.1\ \omega_{\text{s}}$
and $\omega_{\text{max}}=0.5\ \omega_{\text{s}}$, with effective
couplings $\eta_{\text{eff}}=0.5$ (solid red and green lines) and
$\eta_{\text{eff}}=2.0$ (magenta and blue). Results (i) and (iii)
are from calculations without the CL counter term, (ii) and (iv) from
calculations including the CL counter term. The dashed lines represent
eigenvalues of a regular 1D Morse po\-ten\-tial (black) and of a
1D Morse potential modified by the CL counter term according to \equationabb
(\ref{eq:CL_Morse_pot_withCT}) (red). \label{fig:morsecl_lowfreq}}
\end{figure}

\begin{figure}[t!]
\includegraphics[width=0.45\textwidth]{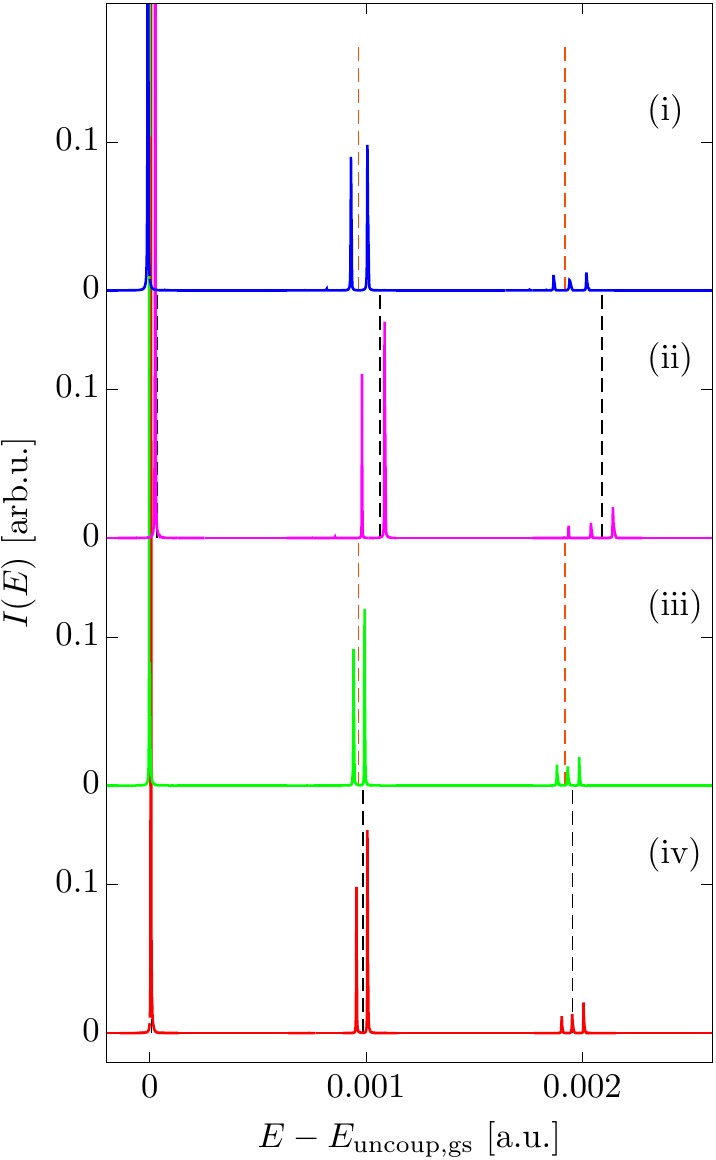}
\caption{Spectra for a Morse oscillator in a CL bath with 20 HOs for high cutoff
frequency. Bath parameters are $\omega_{\text{c}}=0.5\ \omega_{\text{s}}$
and $\omega_{\text{max}}=\omega_{\text{s}}$, with effective couplings
$\eta_{\text{eff}}=0.1$ (red and green lines) and $\eta_{\text{eff}}=0.5$
(magenta and blue). All remaining plot specifications as in \figureabb
\ref{fig:morsecl_lowfreq}. \label{fig:morsecl_highfreq}}
\end{figure}

An exemplary overview of results for the different bath parameters
is given in \figuresabb \ref{fig:morsecl_lowfreq} and \ref{fig:morsecl_highfreq}
for a bath with 20 DOFs, where again all spectra are normalized such
that the respective most intense peak's size is one. To illustrate
the shift of the Morse spectrum, we plot the analytical eigenvalues
of the Morse potential from \equationabb (\ref{eq:CL_Morse_pot_solution})
for calculations without the CL counter term (see red dashed lines
in \figuresabb \ref{fig:morsecl_lowfreq} and \ref{fig:morsecl_highfreq}).
For the evaluation of the calculations with the original CL Hamiltonian
in \equationabb (\ref{eq:CL_Hamiltonian}), the one-dimensional reference
is modified because the counter term, which does not depend on the
bath coordinates, effectively amounts to a renormalization of the
system potential\cite{Costa2000}. Therefore, in this case we use
eigenenergies $E_{\text{mod},n}$ (black dashed lines) of the Morse
potential modified by the CL counter term, 
\begin{align}
V_{\text{s},\text{mod}}(s)=D_{\text{e}}\left(1-\text{e}^{-\alpha\left(s-s_{\text{eq}}\right)}\right)^{2}+\frac{\pi}{4}\frac{F_{\text{b}}}{a}\left(s-s_{\text{eq}}\right)^{2}.\label{eq:CL_Morse_pot_withCT}
\end{align}

The spectra with low bath cutoff frequencies in \figureabb \ref{fig:morsecl_lowfreq}
exhibit system peaks that are hardly different from the 1D result.
If the CL counter term is included (spectra (ii) and (iv)), we see
different blueshifts that can be attributed to the modification of
the system potential by the counter term. For the cases without counter
term (spectra (i) and (iii)), even the difference in effective coupling
strength has little effect, at least on the scale of this figure.
The higher cutoff frequency, on the other hand, has a much greater
impact on the spectra, as depicted in \figureabb \ref{fig:morsecl_highfreq}.
Instead of just the excited states of the system, the original peak
has been split, corresponding to redshifted bath and blueshifted system
excitations as discussed in the previous section. Again, the respective
rightmost peak of each group belongs to the system excitation, while
the others are first and second excited state of the resonant bath
mode. The appearance of these bath peaks, or, more to the point, the
fact that they are no longer suppressed but show up so prominently
here, is due to the fact that the resonant bath mode can be driven
much more effectively by the system than the non-resonant one from
the low-cutoff example. In addition, the resonant HO is now incorporated
into the HK part of the calculation, which does not suppress bath
overtones. The more interesting and more relevant feature for us,
however, is that the stronger system-bath interaction results in a
sizable blueshift of the system both for calculations with and without
CL counter term and always relative to the respective modified or
unmodified one-dimensional eigenvalues. For low effective system-bath
coupling, the difference between the results with and without counter
term is not very pronounced (lower graphs in \figureabb \ref{fig:morsecl_highfreq}),
which seems justified given that the effect of the potential renormalization
is almost insignificant. The high coupling case, on the other hand,
exhibits a greater blueshift of the system if the counter term is
not included. Comparing high and low effective coupling, we see that
an increase of $\eta_{\text{eff}}$ leads to a bigger distance between
system and bath peaks of the same group as a consequence of an enhancement
of the respective trend towards blueshift or redshift.

\begin{figure}[ht!]
\centering \includegraphics[width=0.99\textwidth]{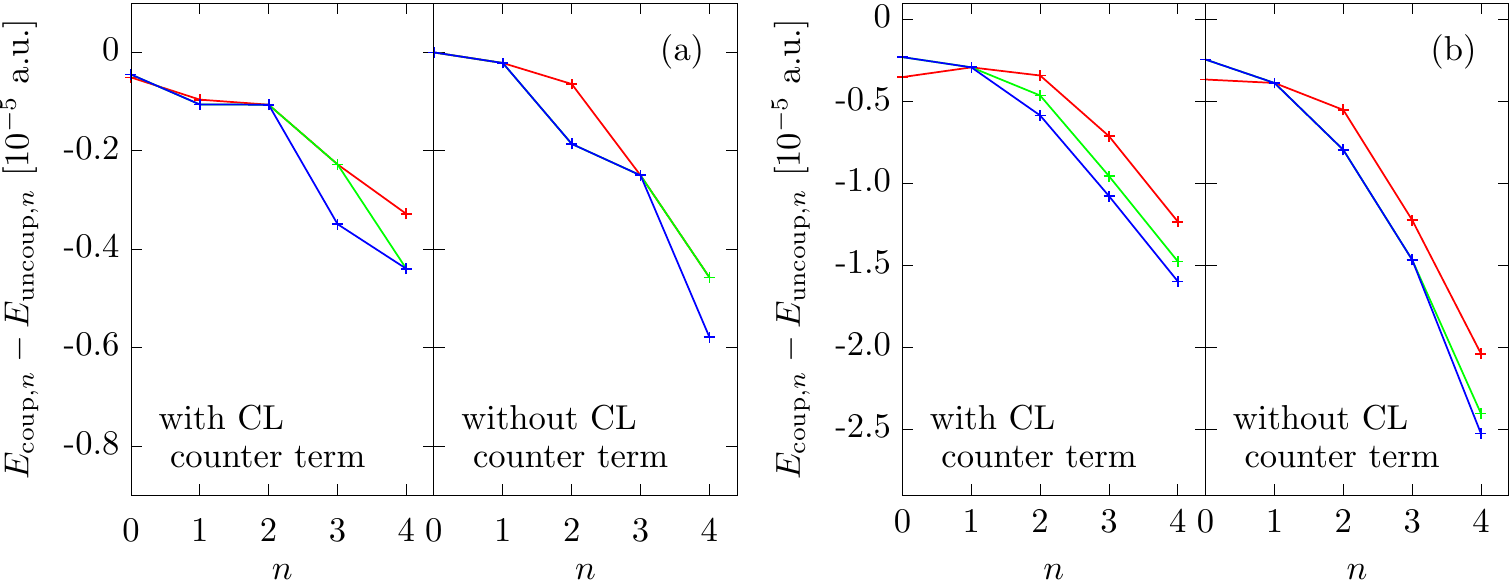}
\caption{Shift of the eigenenergies of a Morse oscillator coupled to a CL bath
with parameters $\omega_{\text{max}}=0.5\ \omega_{\text{s}}$, $\omega_{\text{c}}=0.1\ \omega_{\text{s}}$,
and different coupling strengths: $\eta_{\text{eff}}=0.5$ in panel
(a) and $\eta_{\text{eff}}=2.0$ in panel (b). The bath comprises
either 10 (red crosses), 20 (green), or 40 (blue) HOs, and the solid
lines are just a guide to the eye. The CL counter term according to
\equationabb (\ref{eq:CL_Hamiltonian}) is included in the calculations
on the left side of each panel, and not included on the right side
(\equationabb (\ref{eq:CL_Hamiltonian_noCT})). As a consequence,
the reference eigenenergies for the 1D MO are different depending
on the presence of the counter term, according to \equationabb (\ref{eq:CL_uncoupled}).
\label{fig:morsecl_diff_lowfreq}}
\vspace{0.3cm}
 \includegraphics[width=0.99\textwidth]{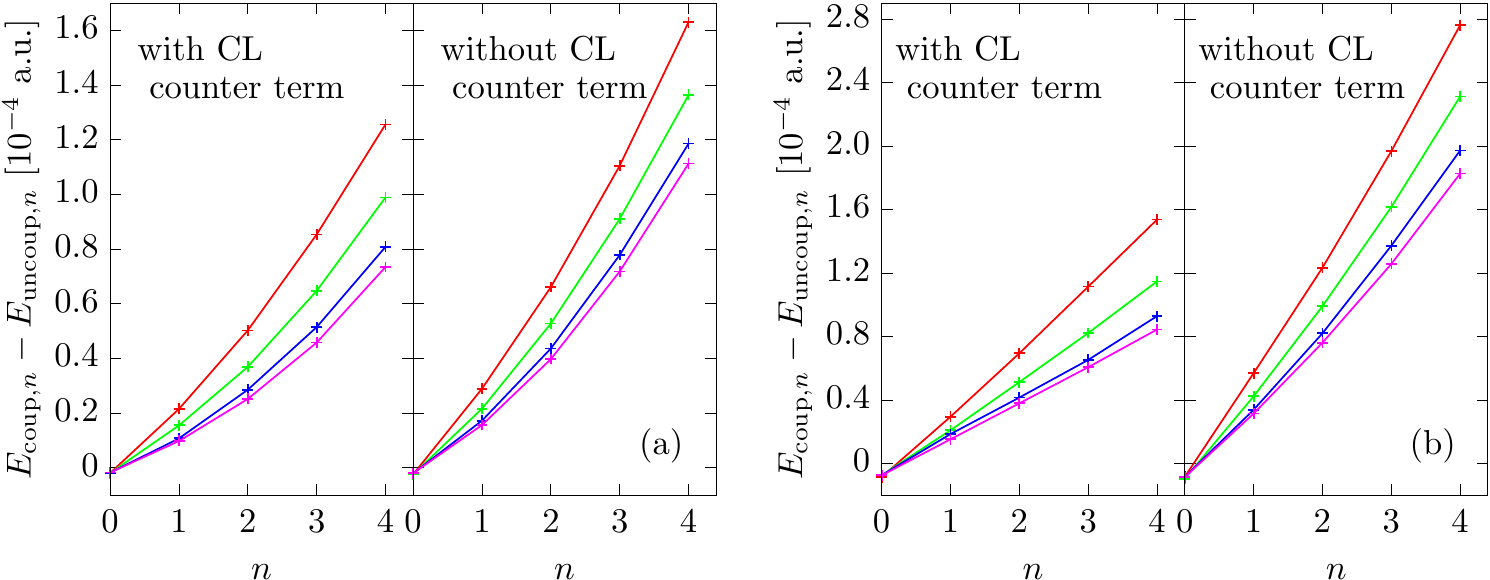} \caption{Shift of the eigenenergies of a MO coupled to a CL bath with parameters
$\omega_{\text{max}}=\omega_{\text{s}}$, $\omega_{\text{c}}=0.5\ \omega_{\text{s}}$,
and coupling strengths $\eta_{\text{eff}}=0.1$ in panel (a) and $\eta_{\text{eff}}=0.5$
in (b). All plot specifications as in \figureabb \ref{fig:morsecl_diff_lowfreq},
with the addition of results for 60 bath DOFs (magenta crosses). \label{fig:morsecl_diff_highfreq}}
\end{figure}

\begin{figure}[th!]
\centering \includegraphics[width=0.99\textwidth]{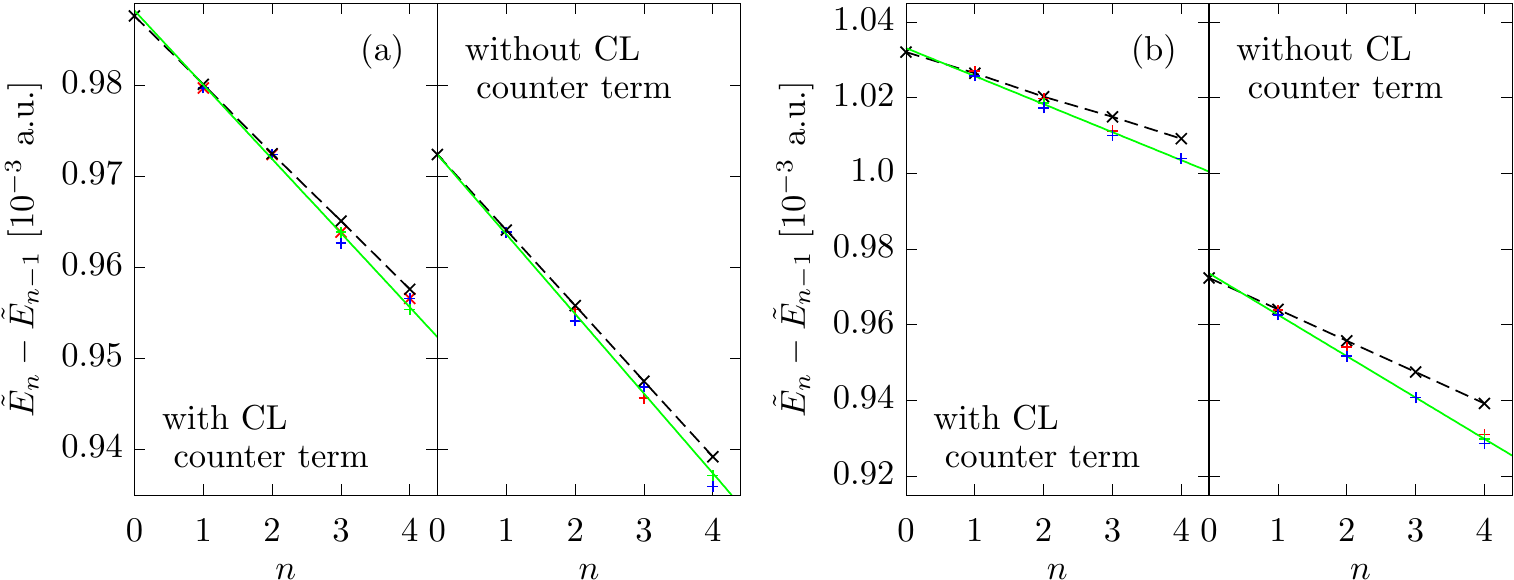}
\caption{Birge-Sponer fit to the difference of consecutive eigenenergies $\tilde{E}_{n}$
of the Morse oscillator coupled to a low-frequency CL bath from \figureabb
\ref{fig:morsecl_diff_lowfreq}. Bath parameters are $\omega_{\text{max}}=0.5\ \omega_{\text{s}}$,
$\omega_{\text{c}}=0.1\ \omega_{\text{s}}$, and couplings are $\eta_{\text{eff}}=0.5$
in panel (a) and $\eta_{\text{eff}}=2.0$ in (b). The ``+'' crosses
denote results with 10 (red), 20 (green), and 40 (blue) bath HOs.
The solid green line is a linear fit to the 20 DOFs result, and the
dashed black line with ``$\times$'' crosses is the corresponding
1D MO reference, i.e., $E_{n}-E_{n-1}$ (\equationabb (\ref{eq:CL_Morse_pot}))
or $E_{\text{mod},n}-E_{\text{mod},n-1}$ (\equationabb (\ref{eq:CL_Morse_pot_withCT})),
respectively.}
\label{fig:morsecl_bisp_lowfreq} \vspace{0.3cm}
 \includegraphics[width=0.99\textwidth]{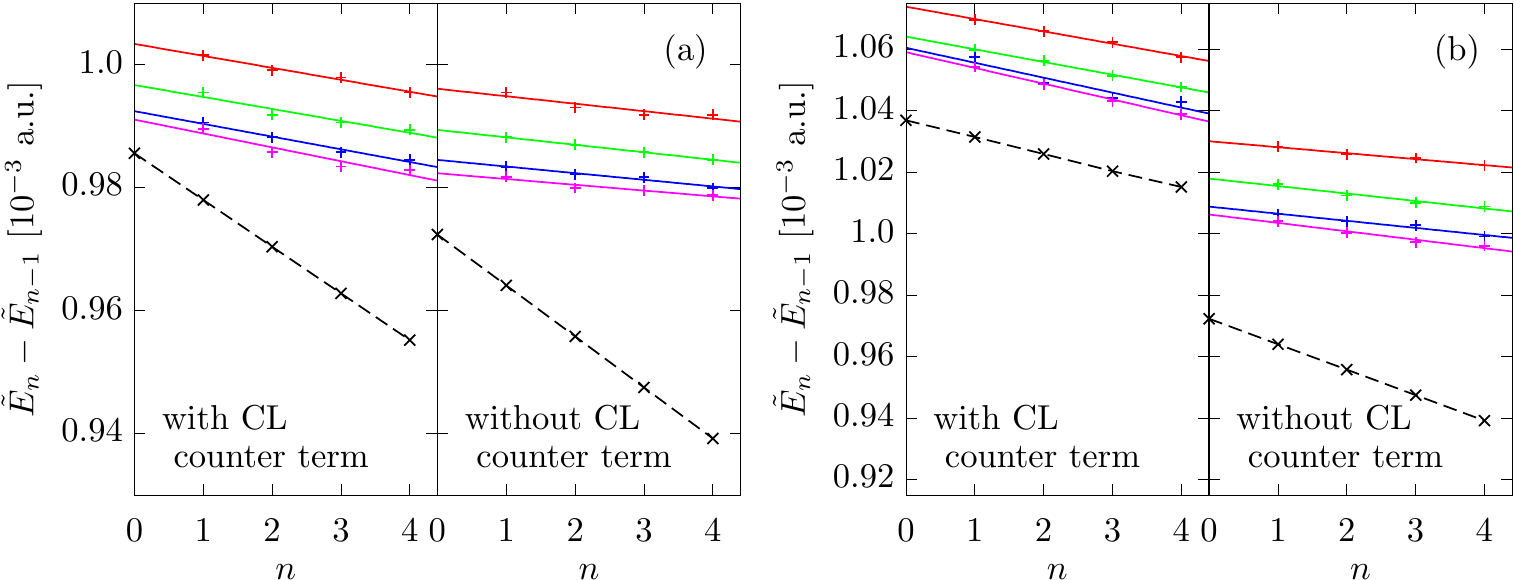} \caption{Birge-Sponer fit to the difference of consecutive eigenenergies $\tilde{E}_{n}$
of the Morse oscillator coupled to a high-frequency CL bath from \figureabb
\ref{fig:morsecl_diff_highfreq}. Bath parameters are $\omega_{\text{max}}=\omega_{\text{s}}$,
$\omega_{\text{c}}=0.5\ \omega_{\text{s}}$, and couplings are $\eta_{\text{eff}}=0.1$
in panel (a) and $\eta_{\text{eff}}=0.5$ in (b). All lines are fits
to points of the same color as in \figureabb \ref{fig:morsecl_bisp_lowfreq},
with the color code from \figureabb \ref{fig:morsecl_diff_highfreq}.}
\label{fig:morsecl_bisp_highfreq} 
\end{figure}

For a more detailed quantitative discussion of the system's blueshift,
we take a look at the first five system peaks for each of the different
baths (\figuresabb \ref{fig:morsecl_diff_lowfreq} to \ref{fig:morsecl_bisp_highfreq}).
In \figuresabb \ref{fig:morsecl_diff_lowfreq} and \ref{fig:morsecl_diff_highfreq},
the shift of the peak energies is plotted. In a similar way as before,
the energy of an appropriate uncoupled reference, now depending on
the peak index, is subtracted, 
\begin{align}
E_{\text{plot},n}=E_{\text{coup},n}-\left(E_{\text{s},n}+\sum_{i=1}^{F_{\text{b}}}\frac{\omega_{i}}{2}\right),\label{eq:CL_uncoupled}
\end{align}
where the system eigenenergies $E_{\text{s},n}$ are either the analytic
eigenenergies $E_{n}$ of the undisturbed Morse potential from \equationabb
(\ref{eq:CL_Morse_pot}) for the calculations without CL counter term,
or the numerically calculated eigenenergies $E_{\text{mod},n}$ of
the modified Morse potential from \equationabb (\ref{eq:CL_Morse_pot_withCT})
for the calculations including the counter term. Thus, we visualize
the net shift of the peaks, which includes the energy shifts of the
system eigenstates and of the bath groundstate. A blueshift of the
system is characterized by a sequence of increasing values, whereas
a redshift shows the opposite behavior. Assuming that the interaction
with the bath only changes the Morse parameters of the system to $\tilde{\omega}_{\text{e}}$
and $\tilde{x}_{\text{e}}$, but not the overall Morse form itself,
\equationabb (\ref{eq:CL_uncoupled}) should have the form of a parabola,
as can be seen by inserting the Morse eigenvalues from \equationabb
(\ref{eq:CL_Morse_pot_solution}), 
\begin{align}
E_{\text{plot},n}=\left(\tilde{\omega}_{\text{e}}-\omega_{\text{e}}\right)\left(n+\frac{1}{2}\right)-\left(\tilde{\omega}_{\text{e}}\tilde{x}_{\text{e}}-\omega_{\text{e}}x_{\text{e}}\right)\left(n+\frac{1}{2}\right)^{2}-\Delta E_{\text{b},\text{gs}}.\label{eq:CL_difference_analytic}
\end{align}
The last term in this equation is the change of the bath ground state
energy upon coupling to the system, which acts as a constant offset.

As an alternative measure, \figuresabb \ref{fig:morsecl_bisp_lowfreq}
and \ref{fig:morsecl_bisp_highfreq} show the difference of consecutive
excited Morse peaks, $\tilde{E}_{n}-\tilde{E}_{n-1}$, 
such that the shift of the bath groundstate energy $\Delta E_{\text{b},\text{gs}}$
drops out. This kind of representation is referred to as Birge-Sponer
extrapolation and can be used experimentally to determine Morse potential
parameters from spectroscopic data \cite{Lewis1994,David2008}. Based
on the analytic formula for the Morse eigenenergies in \equationabb
(\ref{eq:CL_Morse_pot_solution}), a linear fit of these points yields
the harmonic approximation frequency $\omega_{\text{e}}$ as the intersection
with the vertical axis and the anharmonicity $\omega_{\text{e}}x_{\text{e}}$,
which is proportional to the slope of the line. An increase of the
slope corresponds to a redshift whereas a decreasing slope means a
bigger difference between eigenvalues and therefore a blueshift. We
show a linear fit of the first four system energy differences and
compare this result to the one-dimensional Morse oscillator or its
modified version (black ``$\times$'', dashed line), for which the
intersection with the vertical axis has also been obtained by Birge-Sponer
fit. The shifts of the experimental parameters with respect to the
gas phase result are summarized in \tableabb \ref{tab:summary_shifts}.

The analysis is interesting especially for the low-frequency bath,
where we could not see much in the overview plot (\figureabb \ref{fig:morsecl_highfreq}).
Results are presented in \figuresabb \ref{fig:morsecl_diff_lowfreq}
and \ref{fig:morsecl_bisp_lowfreq} for calculations with 10 (red
crosses), 20 (green crosses), and 40 bath DOFs (blue crosses), and
with two different system-bath couplings, $\eta_{\text{eff}}=0.5$
on the left (panel (a)), and for $\eta_{\text{eff}}=2.0$ on the right
(panel (b)). Due to the fits in the Birge-Sponer plots in \figureabb
\ref{fig:morsecl_bisp_lowfreq} being almost identical, we have plotted
only one line in each case. For low coupling, we see an almost negligible
redshift. The effect of the CL counter term is nicely illustrated
by the Birge-Sponer plot: the result with counter term is clearly
blueshifted with respect to the original Morse eigenvalues, but it
is almost on top of the appropriately modified 1D energies. The influence
of the system-bath dynamics is much smaller by comparison, especially
given that our energy grid resolution is $\Delta E=1.2\times10^{-6}\ \text{a.u.}$
These findings are corroborated by the calculations with higher system-bath
coupling strength. Here, the redshift is much more pronounced, but
again, for the original CL potential the main contribution to the
energy shift is due to the counter term. For all calculations with
low bath cutoff frequency, the number of bath oscillators does not
have much impact on the system spectrum. In the low coupling case,
the difference between all three bath sizes is one frequency grid
point at most. For the higher coupling, 10 bath DOFs influence the
system somewhat less than 20 and 40 bath HOs, which yield very similar
results. This weak dependence on bath size is of course a consequence
of the low cutoff and maximum bath frequencies. While the bath mode
with highest frequency is always the same, most additional bath oscillators
are far off-resonant. As a conclusion, we can say that we find the
same 20 to 40 bath DOFs to be sufficient to describe a continuous
bath in this low frequency case, as it has been reported in Ref.\ \onlinecite{Wang2001}.

The case of baths with a high cutoff and resonant maximum frequency
is investigated in detail in \figuresabb \ref{fig:morsecl_diff_highfreq}
and \ref{fig:morsecl_bisp_highfreq}. We have used the same color
scheme for bath size and reference states, but added calculations
with 60 bath DOFs (magenta crosses). The overall behavior of the system
is completely different compared to the low-frequency case, with strong
blueshifts for each bath setup, as already seen in the overview figure
\ref{fig:morsecl_highfreq}. For $\eta_{\text{eff}}=0.1$, the effect
of the bath on the system is somewhat bigger without the CL counter
term, as shown on the left side of \figureabb \ref{fig:morsecl_diff_highfreq}(a).
The counter term, which is harmonic in the system coordinate, restricts
the system dynamics and thus also the system-bath interaction. If
it is left out of the calculation, the system-bath dynamics induces
a larger shift of the system frequency. The total blueshift in \figureabb
\ref{fig:morsecl_bisp_highfreq}, on the other hand, is also determined
by the change of the 1D eigenenergies by the counter term, which offsets
the weaker system-bath dynamics and leads to quite similar overall
results in this low-coupling case. Unlike before, the results strongly
depend on the bath size, which is quite understandable given that
each increase adds in particular some oscillators that are close to
the system frequency and notably influence the system's dynamics.
As the differences between results get smaller with each addition
of bath HOs, we are approaching convergence with respect to a continuous
bath description with 60 bath DOFs. The higher system-bath coupling
amplifies these trends. Now, the difference of coupled and uncoupled
peak energies (\figureabb \ref{fig:morsecl_diff_highfreq}(b)) is
almost twice as big for the calculation without CL counter term compared
to the one that includes it. However, the modification of the 1D eigenenergies
induced by the counter term is so big in this case that the total
blueshift (\figureabb \ref{fig:morsecl_bisp_highfreq}(b)) becomes
even larger than without counter term. Again, we can see that the
results converge with increasing bath size. 
\begin{table}[t]
\centering %
\begin{tabular}{cccc}
\hline 
 & bath w/o resonant modes &  & bath w/ resonant modes\tabularnewline
\hline 
with CL counter term & $\tilde{\omega}_{\text{e}}>\omega_{\text{e}}$  &  & $\tilde{\omega}_{\text{e}}>\omega_{\text{e}}$ \tabularnewline
(\equationabb (\ref{eq:CL_Hamiltonian}))  & $\tilde{\omega}_{\text{e}}\tilde{x}_{\text{e}}<\omega_{\text{e}}x_{\text{e}}$  &  & $\tilde{\omega}_{\text{e}}\tilde{x}_{\text{e}}<\omega_{\text{e}}x_{\text{e}}$ \tabularnewline
\hline 
without CL counter term  & $\tilde{\omega}_{\text{e}}>\omega_{\text{e}}$  &  & $\tilde{\omega}_{\text{e}}>\omega_{\text{e}}$ \tabularnewline
(\equationabb (\ref{eq:CL_Hamiltonian_noCT}))  & $\tilde{\omega}_{\text{e}}\tilde{x}_{\text{e}}>\omega_{\text{e}}x_{\text{e}}$  &  & $\tilde{\omega}_{\text{e}}\tilde{x}_{\text{e}}<\omega_{\text{e}}x_{\text{e}}$ \tabularnewline
\hline 
\end{tabular}\caption{\label{tab:summary_shifts} Change of the harmonic frequency and the
anharmonicity of a Morse oscillator in the presence of a Caldeira-Leggett
bath. All shifts are relative to the gas phase result.}
\end{table}

\begin{figure}[t]
\centering \includegraphics[width=0.49\textwidth]{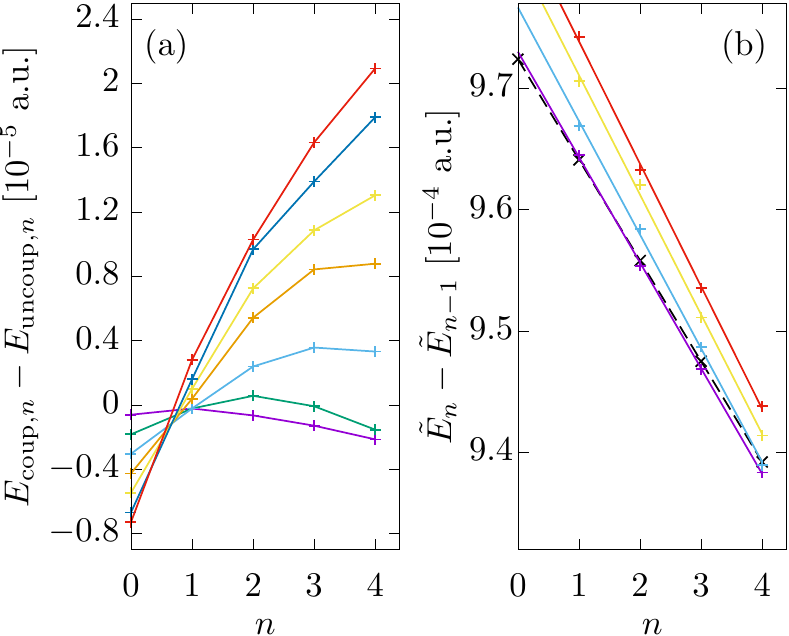}
\caption{Cutoff frequency dependence of the shift of the eigenenergies of a
Morse oscillator coupled to a CL bath. Bath parameters: 20 HOs, maximum
frequency fixed at $\omega_{\text{max}}=0.7\ \omega_{\text{s}}$,
and cutoff frequencies $\omega_{\text{c}}$ are $0.1\ \omega_{\text{s}}$
(violet ``+'' crosses), $0.2\ \omega_{\text{s}}$ (green), $0.3\ \omega_{\text{s}}$
(light blue), $0.4\ \omega_{\text{s}}$ (orange), $0.5\ \omega_{\text{s}}$
(yellow), $0.6\ \omega_{\text{s}}$ (dark blue), and $0.7\ \omega_{\text{s}}$
(red). The CL counter term is not included in the Hamiltonian (\equationabb
(\ref{eq:CL_Hamiltonian_noCT})). The ``$\times$'' crosses and
black dashed line show the eigenenergy differences of the gas-phase
Morse oscillator. The lines in subfigure (a) are just a guide to the
eye. In subfigure (b), the lines are the linear fit to the crosses
of the same color. \label{fig:morsecl_diff_omegac}}
\end{figure}

\begin{figure}[t]
\centering \includegraphics[width=0.49\textwidth]{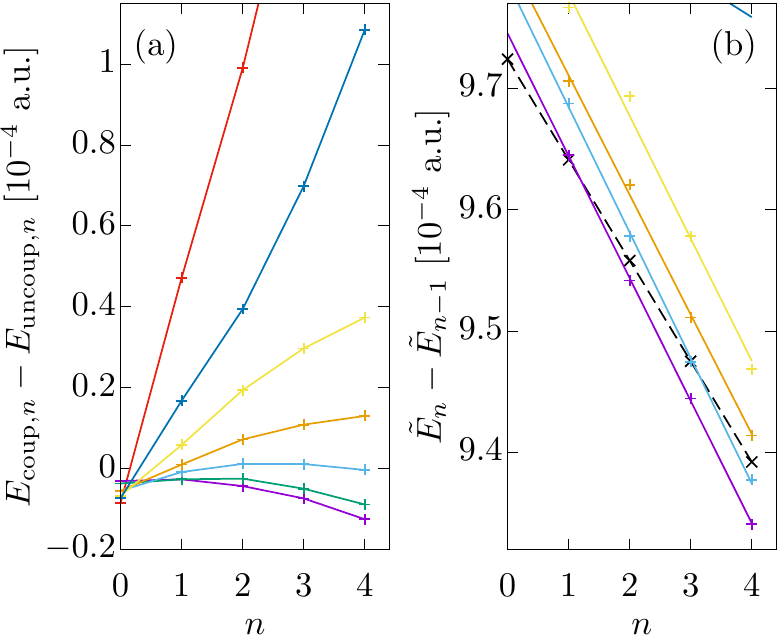}
\caption{Maximum bath frequency dependence of the shift of the eigenenergies
of a Morse oscillator coupled to a CL bath. Bath parameters: 20 HOs,
cutoff frequency fixed at $\omega_{\text{c}}=0.5\ \omega_{\text{s}}$,
and maximum frequencies $\omega_{\text{max}}$ are $0.4\ \omega_{\text{s}}$
(violet ``+'' crosses), $0.5\ \omega_{\text{s}}$ (green), $0.6\ \omega_{\text{s}}$
(light blue), $0.7\ \omega_{\text{s}}$ (orange), $0.8\ \omega_{\text{s}}$
(yellow), $0.9\ \omega_{\text{s}}$ (dark blue), and $1.0\ \omega_{\text{s}}$
(red). The CL counter term is not included in the Hamiltonian (\equationabb
(\ref{eq:CL_Hamiltonian_noCT})). The ``$\times$'' crosses and
black dashed line show the eigenenergy differences of the gas-phase
Morse oscillator. The lines in subfigure (a) are just a guide to the
eye. In subfigure (b), the lines are the linear fit to the crosses
of the same color. \label{fig:morsecl_diff_omegamax}}
\end{figure}

Given the different nature of the system frequency shift in \figuresabb
\ref{fig:morsecl_diff_lowfreq} and \ref{fig:morsecl_diff_highfreq},
one may wonder how the transition from redshift to blueshift looks
like for varying bath cutoff or maximum frequency. This is illustrated
exemplarily in \figuresabb \ref{fig:morsecl_diff_omegac} and \ref{fig:morsecl_diff_omegamax}
for a bath with 20 HOs and using the Hamiltonian without the CL counter
term. In \figureabb \ref{fig:morsecl_diff_omegac}, we keep the maximum
frequency fixed at $\omega_{\text{max}}=0.7\ \omega_{\text{s}}$,
while the cutoff frequency varies between $\omega_{\text{c}}=0.1\ \omega_{\text{s}}$
and $\omega_{\text{c}}=0.7\ \omega_{\text{s}}$. As expected from
the above investigations, the blueshift of the higher eigenfrequencies
is gradually diminished and finally turns into a redshift as the bath
oscillators become more off-resonant. Taking the parabola from \equationabb
(\ref{eq:CL_difference_analytic}) as an appropriate description for
the curves in subfigure (a), we see that indeed all of these graphs
display a negative curvature, which is equivalent to an increase and
therefore a redshift of the anharmonicity $\tilde{\omega}_{\text{e}}\tilde{x}_{\text{e}}$.
This is corroborated by the Birge-Sponer plot in subfigure (b), where
it shows as a steeper slope of the linear fit. The harmonic approximation
frequency $\tilde{\omega}_{\text{e}}$ is always blueshifted, and
the shift is enhanced as $\omega_{\text{c}}$ grows. 

Similar behavior is observed for the opposite case displayed in \figureabb
\ref{fig:morsecl_diff_omegamax}, where we keep the cutoff frequency
fixed at $\omega_{\text{c}}=0.5\ \omega_{\text{s}}$ and vary the
maximum frequency between $\omega_{\text{max}}=0.4\ \omega_{\text{s}}$
and $\omega_{\text{max}}=1.0\ \omega_{\text{s}}$. The parabolas in
subfigure (a) now change from negative to positive curvature as the
the maximum bath frequency becomes resonant. This indicates a blueshift
of the anharmonicity if there are close to resonant modes present
in the bath, as already seen in \figuresabb \ref{fig:morsecl_diff_highfreq}
and \ref{fig:morsecl_bisp_highfreq}. As the maximum frequency becomes
more off-resonant, we see again a redshift of the anharmonicity. In
both cases, the harmonic frequency is shifted to a higher value.

\section{Conclusions and Outlook}

\label{sec:conc}

We have shown that the time averaged mixed semiclassical hybrid approach
introduced recently \cite{Buchholz2016} can be applied to study spectroscopic
signatures of an anharmonic system of interest in a large environment,
with up to 60 bath degrees of freedom. Conditions on the bath parameters
have been identified, that lead either to a redshift or to a blueshift
of the system frequency. We have investigated the cases of a non-resonant
and a resonant bath where in the latter case also the bath oscillator
closest to resonance was treated on the full HK level. Furthermore,
we have compared results from calculations with and without the Caldeira-Leggett
counter term to demonstrate that this term causes a large portion
of the blueshift which is observed with respect to the gas phase Morse
oscillator. If the one-dimensional reference potential is adjusted
appropriately with the counter term, the effect of the system-bath
interaction on the system eigenenergies is similar for calculations
with and without counter term. The change of the system frequency
depends both on the bath cutoff and the maximum frequency. In the
case of a strongly non-resonant bath, the anharmonicity $\omega_{\text{e}}x_{\text{e}}$
is redshifted. The harmonic frequency $\omega_{\text{e}}$ always
shifts to a higher value. If there are at least some bath frequencies
close to resonance, on the other hand, the fundamental frequency as
well as the anharmonicity are blueshifted. Overall, the mixed time
averaged semiclassical hybrid approach demonstrated to be a robust
semiclassical approximation that properly accounts for different types
of coupling, even for up to 61-dimensional potential. This opens the
route of its application to more realistic condensed phase systems
than the Caldeira-Leggett potential modeling.

Recently, it has been argued that the modeling of an anharmonic system
bilinearly coupled to a harmonic bath suffers the invertibility problem
\cite{Gottwald2015}. The next goal that we intend to tackle therefore
is the study of realistic system bath Hamiltonians with Lennard-Jones
type interaction, for systems like iodine in a Krypton matrix, which
have also been studied experimentally \cite{Karavitis2001}.

Future implementation will include the finite temperature effects
and the broadening of the peaks induced by the temperature. Also different
types of semiclassical approximation, such as Linearized (LSC-IVR)
and van Vleck SC-IVR (VV-SC-IVR), can be equally mixed as it was done
for the time-averaging SC-IVR and the TGWD. Eventually,
given the cheap computational cost of the time-averaging SC-IVR, the
present mixed semiclassical method will be implemented for direct
\emph{ab initio} simulations of condensed phase systems.

\section{Acknowledgements}

Michele Ceotto and Max Buchholz acknowledge financial support from
the European Research Council (ERC) under the European Union\textquoteright s
Horizon 2020 research and innovation programme (Grant Agreement No.
{[}647107{]} \textemdash{} SEMICOMPLEX \textemdash{} ERC-2014-CoG).
M.C. acknowledges also the CINECA and the Regione Lombardia award
under the LISA initiative (grant GREENTI) for the availability of
high performance computing resources. M. B. acknowledges funding from
the Graduate Academy of TU Dresden as well as from the IMPRS of the
MPI-PKS Dresden.

\newpage{}

\bibliographystyle{apsrev4-1}
\bibliography{literature_submitted}

\end{document}